\documentclass[fleqn,usenatbib]{mnras}

\usepackage{newtxtext,newtxmath}
\usepackage[T1]{fontenc}
\usepackage{ae,aecompl}
\usepackage{graphicx}   
\usepackage{amsmath}    
\usepackage{amssymb}   

\title[HODs for accurate clustering predictions]{Extensions to the halo occupation distribution model for more accurate clustering predictions}

\author[Esteban Jim\'{e}nez et al.]{
Esteban Jim\'{e}nez $^{1}$\thanks{E-mail: enjimenez@uc.cl},
Sergio Contreras$^{2}$,
Nelson Padilla$^{1,3}$,
Idit Zehavi$^{4}$,
\newauthor
Carlton M. Baugh$^{5}$ and
Violeta Gonzalez-Perez $^{6,7}$
\\
$^{1}$Instituto de Astrof\'{i}sica, Pontificia Universidad Cat\'{o}lica de Chile, Santiago, Chile\\
$^{2}$Donostia International Physics Center (DIPC), Manuel Lardizabal pasealekua 4, 20018 Donostia, Basque Country, Spain\\
$^{3}$Centro de Astro-Ingenier\'{i}a, Pontificia Universidad Cat\'{o}lica de Chile, Santiago, Chile\\
$^{4}$Department of Physics, Case Western Reserve University, Cleveland, OH 44106, USA\\
$^{5}$Institute for Computational Cosmology, Department of Physics, Durham University, South Road, Durham, DH1 3LE, UK\\
$^{6}$Institute of Cosmology \& Gravitation, University of Portsmouth, Dennis Sciama Building, Portsmouth, PO1 3FX, UK.\\
$^{7}$Energy Lancaster, Lancaster University, Lancaster LA14YB, UK
}

\date{Accepted XXX. Received YYY; in original form ZZZ}
\pubyear{2019}

\begin{document}
\label{firstpage}
\pagerange{\pageref{firstpage}--\pageref{lastpage}}
\maketitle

\begin{abstract}
We test different implementations of the halo occupation distribution (HOD)
model to reconstruct the spatial distribution of galaxies as predicted by 
a publicly available semi-analytical model (SAM). We compare the measured two-point 
correlation functions of the HOD mock catalogues and the SAM samples to quantify the
fidelity of the reconstruction. We use fixed number density galaxy samples
selected according to stellar mass or star formation rate (SFR). 
We develop three different schemes to populate haloes with galaxies with increasing complexity, considering the scatter of the satellite
HOD as an additional parameter in the modelling. We first modify the SAM output, removing
assembly bias and using a standard Navarro-Frenk-White density profile for the satellite
galaxies as the target to reproduce with our HOD mocks. We find that all models give similar reproductions of the two-halo 
contribution to the clustering signal, but there are differences in the one-halo term. 
In particular, the HOD mock reproductions work equally well using either the HOD of 
central and satellites separately or using a model that also accounts for whether or not 
the haloes contain a central galaxy. We find that the HOD scatter does not have an important
impact on the clustering predictions for stellar mass selected samples. For SFR selections, 
we obtain the most accurate results assuming a negative binomial distribution for the number of satellites in a halo.
The scatter in the satellites HOD is a key consideration for HOD mock catalogues that mimic
ELG or SFR selected samples in future galaxy surveys.
\end{abstract}


%%%%%%%%%%%%%%%%%%%%%%%%%%%%%%%%%%%%%%%%%%%%%%%%%%%%%%%%%%%%%%%%%%%%%%%%%%%%%%%%%%%%%%%%
%                                     INTRODUCTION
%%%%%%%%%%%%%%%%%%%%%%%%%%%%%%%%%%%%%%%%%%%%%%%%%%%%%%%%%%%%%%%%%%%%%%%%%%%%%%%%%%%%%%%%
\section{Introduction}

In the current cosmological paradigm, the Universe is composed of a filamentary network of structures shaped by gravity. In this framework, dark matter haloes correspond to overdense regions that evolve by gravitational instability due to mergers and interactions with other haloes. Galaxy formation occurs inside haloes where baryons collapse in the gravitational potentials and the condensation of cold gas allows the formation of stars and the evolution of galaxies \citep{White78}. A detailed description of the halo-galaxy connection enables using the galaxies to constrain the cosmological model. 

The evolution of dark matter haloes can be followed, to high accuracy, using N-body simulations which use a set of cosmological parameters as inputs. In contrast, the evolution of galaxies in  haloes involves many physical processes that are still poorly understood. The fate of baryons within dark matter haloes has been modelled using different approaches, such as hydrodynamical simulations that provide an insight into the formation and evolution of galaxies \citep[e.g.][]{Vogelsberger14,Schaye15}. However, these models are computationally expensive and cannot be run over the large volumes needed for cosmological studies. Alternatively, the effect of baryons can be probed in such large volumes using semi-analytical models (SAMs) of galaxy formation. These start from haloes extracted from a large volume dark matter only simulation and use simplified physical models of the processes that shape the evolution of baryons \citep{cole2000,Baugh06-rv,Benson10,Sommerville15}. Hence, SAMs make predictions for the abundance and clustering of galaxies that can be compared and tested with large surveys. 

Another way to describe the galaxy population is with the halo occupation distribution (HOD) framework \citep{Benson00, Peacokc00,Scoccimarro01,Yang03}. This is an empirical approach that provides a relation between the mass of haloes and the number of galaxies hosted by them. This is expressed as the probability distribution $P(N|M_h)$ that a halo of virial mass $M_h$ hosts $N$ galaxies which satisfy some selection criteria. This approach provides insight into the halo-galaxy connection and can be used to study galaxy clustering \citep{Berlind02,Zheng05,Conroy06,Zehavi11,Wechsler18}. Furthermore, the HOD parameters can be tuned in detail because they only aim to reproduce a limited set of observables such as the galaxy number density and clustering. Thus, HOD modelling is one of the most efficient ways to populate very large volumes or to produce many realizations required for, e.g., estimating covariance matrices using mock galaxy catalogues \citep[e.g.][]{Norberg:2009, Manera13}. These mock catalogues can then be used to test and develop new algorithms that will be used for the next generation of surveys. 

The study of star forming emission line galaxies (ELGs) has gained interest over the last decade as they will be targeted by surveys such as Euclid and the Dark Energy Spectroscopic Instrument (DESI) surveys \citep{Laureijs:2011, DESI16}. The luminosity of an emission line depends on a number of factors, including the star formation rate (SFR), gas metallicity and the conditions in the HII regions \citep[e.g.][]{Orsi14}. Even though ELGs samples are related to star formation, they are not the same as SFR-selected samples. Still, a similar HOD approach can be used to study both galaxy populations \citep{Geach12, Cochrane17, Cochrane18}. In particular, the shape of the HOD in SFR selected samples is more complex than the case of the more widely studied stellar mass selected samples \citep[e.g.][]{contreras13,gp18}. For example, the occupation function of central galaxies in ELG samples does not follow the canonical step-like form. Accurate modelling of the HOD will provide the more realistic mock catalogues needed for the analysis of future observational samples. 

Here, we use the HOD formalism to test three different ways to populate dark matter haloes with galaxies. The prescriptions of these models aim to replicate as accurately as possible the target galaxy populations of a SAM sample. The comparison between the galaxy population in the mock catalogues and SAM samples is done via the analysis of their two-point correlation function (2PCF), which is related to the power spectra of density fluctuations and is sensitive to cosmology \citep[e.g.][]{DeRose19}. We also include the scatter of the HOD of satellites in our modelling, and quantify the impact of using this additional parameter on the clustering.

The outline of this paper is as follows. The definition of galaxy samples used and the basic properties of the N-body simulation and the SAM are given in Section~\ref{sec: simulation data}. The correlation functions and the HODs of the samples are presented in Section~\ref{sec: Characterization of samples}. In Section~\ref{sec: building a mock} we introduce the HOD models used to build the mock catalogues and the recipes employed to perform this procedure. The main results and analysis are discussed in Section~\ref{sec: results} while in Section~\ref{sec: conclusions} we present our conclusions. Appendix~\ref{section:1-HODoccupationfunctions} shows the predicted occupation functions for a particular HOD model.
%%%%%%%%%%%%%%%%%%%%%%%%%%%%%%%%%%%%%%%%%%%%%%%%%%%%%%%%%%%%%%%%%%%%%%%%%%%%%%%%%%%%%%%%
%                                     SIMULATION DATA
%%%%%%%%%%%%%%%%%%%%%%%%%%%%%%%%%%%%%%%%%%%%%%%%%%%%%%%%%%%%%%%%%%%%%%%%%%%%%%%%%%%%%%%%
\section{Simulation data} 
\label{sec: simulation data}

In this section we give a brief overview of the galaxy formation model used (\S~2.1) and the N-body simulation in which it is implemented (\S~2.2).

\subsection{Galaxy formation model}

A galaxy formation model needs to take into account a variety of physical processes such as radiative cooling of gas; AGN, supernovae and photoionisation feedback; chemical evolution; star formation; disc instabilities; collapse and merging of dark matter haloes; and galaxy mergers. These affect the fate of baryons in haloes which lead to the formation and evolution of galaxies. Several physical processes such as star formation and gas cooling are not fully understood due to their complexity. As a consequence, a set of free parameters are used in the equations that model these processes. These free parameters are tuned in order to reproduce observations such as the luminosity functions, colours and the distribution of morphological types. In this context, different SAMs usually have their own implementations to model these physical processes, predicting different galaxy populations. Here we use the outputs at $z=0$ from the SAM of \citet{Guo13} (hereafter G13) which is a version of the {\tt  L-GALAXIES} code from the Munich group \citep{DeLucia04, Croton06, DeLucia07, guo11, Henriques13}. The outputs are publicly available from the Millennium Archive\footnote{\url{http://gavo.mpa-garching.mpg.de/Millennium/}}.

The samples used here are defined according to three different number densities where we rank the galaxies in the SAM by their stellar mass or SFR in a decreasing way (hereafter the SAM samples). These samples are useful in order to compare with observational catalogues with similar space densities. Table~\ref{tab: density samples} shows the three number densities and the cuts in stellar mass and SFR used in each case. 

\subsection{The Millennium simulation}
\label{sec:simulation data}

The distribution of dark matter haloes used in this work is drawn from the Millennium-WMAP7 simulation \citet{Guo13} which is identical with the Millennium Simulation \citet{Springel05}, but with updated cosmological parameters that match the results from the WMAP7 observations. This version assumes a flat $\Lambda$CDM universe considering $\Omega_m = 0.27$, $\Omega_{\Lambda} = 0.73$, $h = H_0/(\rm 100\ km\ s^{-1}\ Mpc^{-1}) = 0.704$ and $\sigma_{8} = 0.81$. The simulation was carried out in a box-size of $500\ h^{-1} \rm Mpc$ following $2160^3$ particles of mass $9.31\times 10^{8}h^{-1}\rm M_{\odot}$.  The run produced 61 simulation snapshots from $z=50$ up to $z=0$.  G13 use a friends-of-friends group finding algorithm ({\tt FOF}) to identify dark matter haloes in each snapshot \citep{Davis85} and then run {\tt SUBFIND} to identify the subhaloes \citep{Springel01}. Halo merger trees are constructed for each output and track the evolution of haloes through cosmic time. These trees are the starting points for the SAM.

\begin{table}
    \centering
    \begin{tabular}{ccc}
        \hline
         $n/h^{3}\rm Mpc^{-3}$ & $ {\rm M_{min}^*}/ h^{-1} {\rm M_{\odot}}$ & $\rm SFR_{min}/yr^{-1}M_{\odot}$ \\
        \hline
        $\rm 10^{-3.0}$ & $5.95\times 10^{10}$ & $5.25$ \\
        $\rm 10^{-2.5}$ & $3.38\times 10^{10}$ & $2.53$ \\
        $\rm 10^{-2.0}$ & $1.25\times 10^{10}$ & $0.70$\\
        \hline
    \end{tabular}
    \caption{The first column shows the abundance of galaxies in the three density samples used here. The second and third columns show the cuts applied to G13 galaxies in stellar mass and star formation rate, respectively, to achieve these abundances.}
    \label{tab: density samples}
\end{table}

%%%%%%%%%%%%%%%%%%%%%%%%%%%%%%%%%%%%%%%%%%%%%%%%%%%%%%%%%%%%%%%%%%%%%%%%%%%%%%%%%%%%%%%%
%                   CHARACTERIZATION OF THE SAM GALAXY SAMPLES
%%%%%%%%%%%%%%%%%%%%%%%%%%%%%%%%%%%%%%%%%%%%%%%%%%%%%%%%%%%%%%%%%%%%%%%%%%%%%%%%%%%%%%%%
\section{Characterization of the SAM galaxy samples}
\label{sec: Characterization of samples}

This section introduces the statistics used to characterize the distribution of galaxies, starting with the measurement of the correlation function (\S~3.1), the form of the HOD predicted by the SAM (\S~3.2) and the scatter in the HOD (\S~3.3).  

\subsection{Clustering measurement: two-point galaxy correlation function}
\label{subsec: xi}

The spatial two-point correlation function, $\xi(r)$,  measures the excess probability of finding a pair of galaxies at a given separation with respect to a random distribution. We compute the 2PCF of the galaxy samples with the {\tt Corrfunc} code \citep{Sinha17}.

\begin{figure}
    \centering
    \includegraphics[width=\columnwidth]{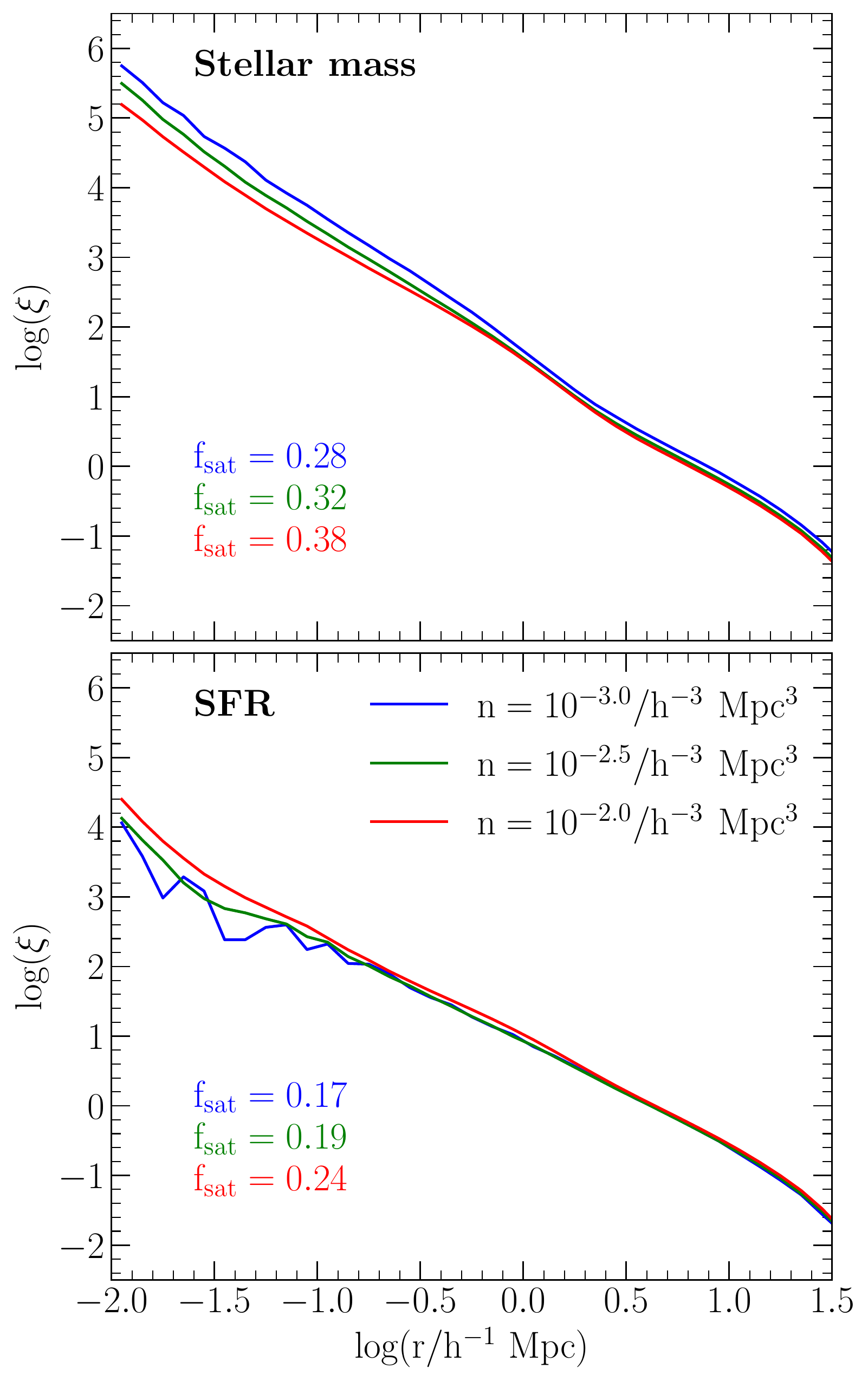}
    \caption{Two-point correlation functions ($\xi(r)$) of different galaxy samples from G13 and defined in Table~\ref{tab: density samples}. ({\it Top}) stellar mass and SFR selected samples ({\it bottom}). Colors indicate each sample as labeled in the bottom panel. The fraction of satellites in each sample is shown in both panels, with the color indicating the  sample number density.} 
    \label{fig: xi from SAMs}
\end{figure}

Fig.~\ref{fig: xi from SAMs} shows the 2PCF of the stellar mass (top) and SFR (bottom) selected samples for the different space densities. For the former, the amplitude of the clustering increases with decreasing number density, as we consider more massive galaxies. The impact of the inclusion of these massive galaxies is stronger at small scales and is weaker at large scales. In contrast, for the SFR selected galaxies the amplitude of the 2PCF for the different samples remains largely unchanged except for small scales where the satellite-satellite pairs make an important contribution to the clustering amplitude. For both selections, the satellite fraction increases with increasing number density.

In the 2PCF, we can distinguish between the contribution from galaxy pairs in the same halo and from different haloes. The former are the main contributors to the amplitude of the 2PCF on small scales, namely the one-halo term which dominates up to $\sim 1\ h^{-1} {\rm Mpc}$, while galaxy pairs between different haloes contribute mostly to the two-halo term which determines the clustering on large scales. 
%In this regime, the amplitude of the 2PCF depends mainly on the number of centrals and the bias of the haloes in which they reside. 
In this regime the total number of galaxies in the halo, regardless of whether they are satellites or the central,  drives the amplitude of the clustering, acting as a weighting for the bias of each halo in computing an overall ``effective'' bias for the sample (see e.g. \citealt{Baugh:1999}).  
The one-halo term is sensitive not only to the number of satellites, but also depends on their spatial distribution.  

%%%%%%%%%%%%%%%%%%%%%%%%%%%%%%%%%%%%%%%%%%%%%%%%%%%%%%%%%%%%%%%%%%%%%%%%%%%%%%%%%%%%

\subsection{The halo occupation function predicted by the SAMs}
\label{subsec: HODs from SAMs}

The galaxy populations in the SAMs depend on the choices adopted for the modelling of the baryonic processes. Hence, depending on the SAM employed, different galaxy catalogues with different luminosity functions, stellar mass functions or correlation functions can be obtained for the same dark matter simulation. For example, \citet{contreras13} studied the effects on the clustering predicted from different SAMs and found some differences particularly in galaxy samples selected by SFR and cold gas mass. Moreover, they show that the shapes of the HODs are model-dependent which reflects  the differences in the implementation of physical processes in each SAM. For example, the specific modelling of dynamical friction affects the satellite population in SAMs. Here we are not interested in the detailed shape of the HOD predicted by a particular SAM, but on how best to use the occupation functions to populate dark matter haloes with galaxies to produce a similar spatial distribution to that resulting from a SAM.

\begin{figure}
    \centering
    \includegraphics[width=\columnwidth]{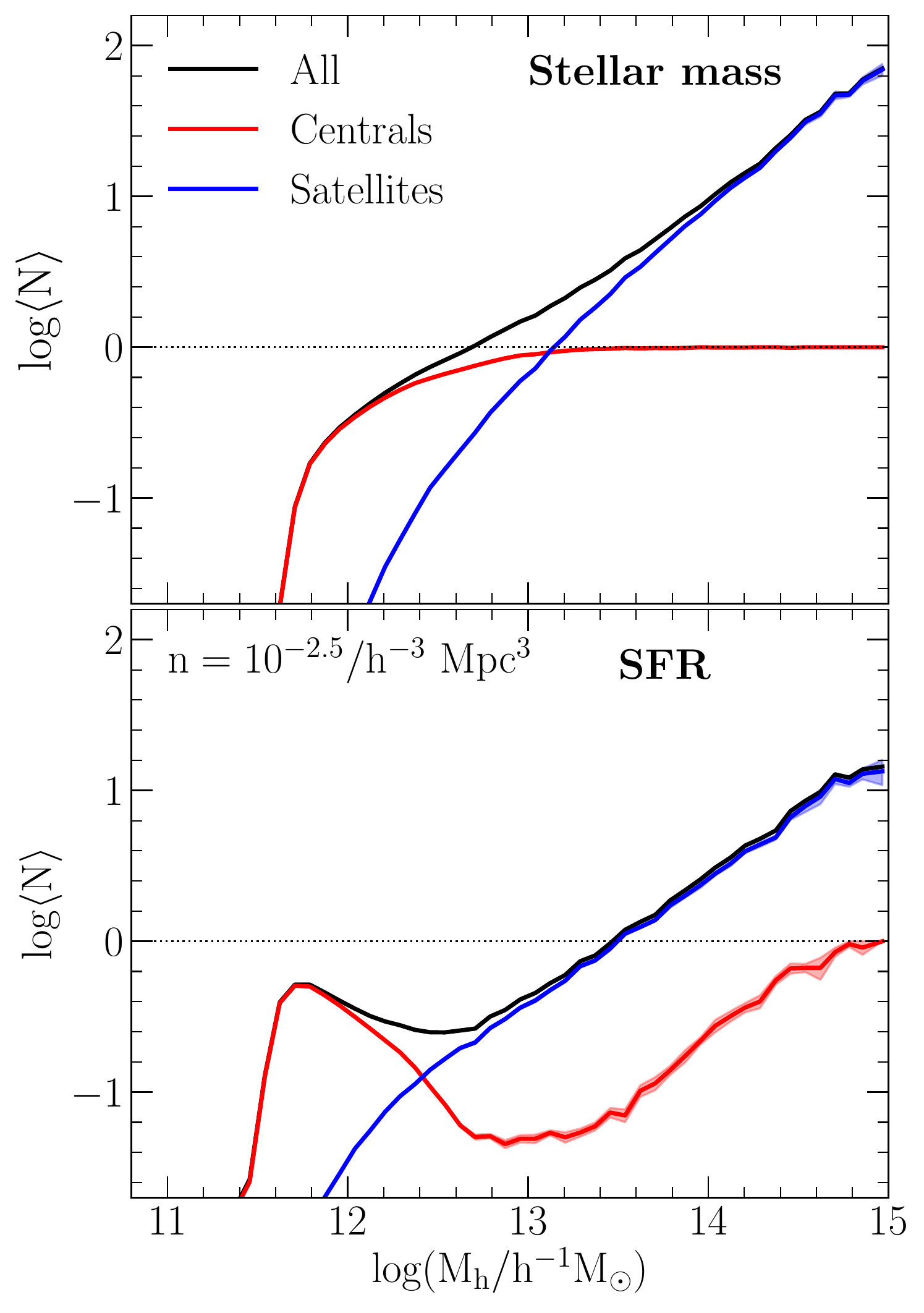}
    \caption{The HOD predicted by G13 for stellar mass ({\it top}) and SFR selected samples ({\it bottom}), for a number density of $10^{-2.5} h^3 \rm Mpc^{-3}$. Black lines show the HOD for the full sample and red and blue indicate the HOD for central and satellite galaxies, respectively.  The red and blue shaded regions represent jackknife errors calculated using 10 subsamples.
    The horizontal black dotted line shows an average occupation value of unity.}
    \label{fig: HODs from SAMs}
\end{figure}

The HOD is usually broken down into the contribution from central and satellite galaxies. Fig.~\ref{fig: HODs from SAMs} shows these two components for stellar mass and SFR selected samples with the same number density for the G13 SAM. Here, each HOD is computed in bins of width 0.08 dex in the logarithm of the halo mass where the position of each $\langle N \rangle$ value is plotted at the median value within each bin. The striking difference in the shape of the HOD of centrals between the two selections is due to the different galaxies that are included. Massive centrals tend to be red galaxies hosted by massive dark haloes. Such centrals are included in stellar mass selected samples but not when selecting by SFR. The galaxies in the SFR samples correspond mainly to blue star-forming galaxies excluding luminous red galaxies with high stellar mass but low SFR. It is noteworthy that the fraction of halos that contain a central passing the SFR selection never reaches unity for the sample plotted in Fig.~\ref{fig: HODs from SAMs}. These features of  the HOD of SFR galaxies have been noted in SAMs before  \citep[e.g.][]{contreras13,contreras19,gp18} and inferred for blue galaxies in the SDSS  \citep[e.g.][]{Zehavi11}. This shows that a significant number of haloes in the SFR selected samples do not host a central as their SFR is below the threshold. The same situation is found in the other number density samples. Note that, in observational samples, the ranking of galaxies in order of their emission line luminosity may not correspond to the ranking in SFR due to dust attenuation, which means that the highest SFR galaxies may not necessarily have the brightest emission lines.  

We estimate the uncertainties of the HOD values using jackknife resampling \citep{Norberg:2009}, dividing the simulation volume into 10 slices. We use the position of the centre of the potential of haloes to classify the galaxies within each halo. The resulting errors are shown as the shaded regions in Fig.~\ref{fig: HODs from SAMs}, and they are negligible for all halo masses except at the high mass end and for the HOD of centrals selected by SFR. 

Because of the simple relation between halo mass and occupation number, the HOD represents a useful approach for the construction of mock galaxy catalogues. Here we have described the first moment of the HOD, the main ingredient to building-mocks recipes. Nevertheless, it is important to also consider the second moment, i.e the dispersion in the HOD of satellites.  

%%%%%%%%%%%%%%%%%%%%%%%%%%%%%%%%%%%%%%%%%%%%%%%%%%%%%%%%%%%%%%%%%%%%%%%%%%%%%
\subsection{The predicted dispersion in the halo occupation number}
\label{subsec: HOD dispersion}

When the simplest HOD approach is used to build mock catalogues, the mean of the distribution is the main parameter. Central galaxies are assumed to follow a nearest-integer distribution where the mean $\rm \langle N_{cen} \rangle$ is between zero and one. For satellites, a Poisson distribution with mean $\rm \langle N_{sat} \rangle$ is the most widely assumed distribution \citep[e.g][]{Kravtsov04,Zheng05}. 

In G13, satellites are classified as type-1 if they are hosted by a resolved subhalo, and type-2 or orphans if the subhalo has been destroyed by tidal effects and is no longer identified. \citet{BK10} found that the number of low mass subhaloes in main haloes in the Millennium-II Simulation \citep{BK09} is well described by a negative binomial distribution which corresponds to a super-Poissonian statistic as its scatter is larger than a Poisson distribution. This suggests that the type-1 satellite population can also be described by this distribution. Based on the outputs from the SAM presented in \citet{Jian16-I}, and using the Bolshoi \citep{Klypin11} and MultiDark \citep{Prada12} simulations, \citet{Jiang16} showed that ignoring this non-Poissonity in the HOD of subhaloes results in systematic errors in the predicted clustering of galaxies. Here we extend the application of the negative binomial distribution by checking whether the HOD of G13 satellites, that is including type-1 and type-2, is well described by this statistic. We expect that the HOD scatter is model-dependent because of the different treatments of dynamical friction. Moreover, as some galaxy properties, such as SFR and stellar mass, have a model-dependent scatter it is reasonable to assume the same for HODs. For example, \citet{Guo16} showed that different galaxy formation models do not have  the same dispersion in the stellar mass-halo mass relation. Therefore our results are specific to the G13 model. It is likely that a different SAM would require an adjustment to to the value of $\beta$ to describe the scatter of the satellite HOD. Nevertheless we expect our general results to hold for any SAM, and that the satellite distribution displays more scatter than Poisson.  

The Poisson and negative binomial distributions differ in their shapes so it is useful to parametrise the departure from the Poisson scatter. We use the parameter $\beta$ (defined below) to denote this departure. For a Poisson distribution the variance is given by the mean value of the random variable, namely $\rm \langle N_{sat}\rangle$, with the standard deviation given by $\sigma = \sqrt{\rm \langle N_{sat}\rangle}$. The negative binomial distribution has the same mean as the Poisson distribution, but a larger scatter which can be expressed as

\begin{equation}
     \sigma_{\rm NB} = \sigma + \beta \sigma.
\end{equation}

\noindent where $0<\beta<1$. Then, $\beta$ indicates the fractional change in the variance with respect to the Poisson standard deviation $\sigma$. Under this definition, when $\beta=0$ the distribution is Poissonian and if $\beta =1$ the standard deviation is twice that from a Poisson distribution.  

The probability function of the negative binomial distribution is given by

\begin{equation}
 P(N\left| r,p \right.) = \frac{\Gamma(N+r)}{\Gamma(r)\Gamma(N+1)}p^r (1-p)^N.
\end{equation}

\noindent Here $\Gamma(x) = (x+1)!$ is the gamma function. The parameters $r$ and $p$ are determined by the first moment $\left< N\right>$ and second moment $\sigma^2$ of the distribution,

\begin{equation} \label{eq. parameters of NB}
 p = \frac{\left< N\right>}{\sigma_{\rm NB}^2}, \hspace{2mm} r = \frac{\left< N\right>^2}{\sigma_{\rm NB}^2 - \left< N\right>}.
\end{equation}

Thus, we can control the width of the negative binomial distribution through the parameter $\beta$ and compute the value of $\sigma_{\rm NB}^2$.

%%%%%%%%%%%%%%%%%%%%%%%%%%%%%%%%%%%%%%%%%%%%%%%%%%%%%%%%%%%%%%%%%%%%%%%%%%%%%%%%%%%%%%%%%%
 %                          GENERATING HOD MOCK CATALOGUES
%%%%%%%%%%%%%%%%%%%%%%%%%%%%%%%%%%%%%%%%%%%%%%%%%%%%%%%%%%%%%%%%%%%%%%%%%%%%%%%%%%%%%%%%%%

\section{Generating HOD mock catalogues}
\label{sec: building a mock}

We now describe the procedure followed to build HOD mock galaxy catalogues using the HODs of the SAM samples.  Section~\ref{subsec: HOD models} presents the three methods we use to populate haloes with galaxies. In Section~{\ref{subsec: scatter in the HOD}} we specify the treatment of the scatter in the HOD of satellites. Section~{\ref{subsec: distribution of satellites}} explains how we impose a standard Navarro-Frenk-White (NFW) density profile for satellites. Section~\ref{subsec: AB and shuffle} presents the impact of assembly bias in the SAM samples and explains why it must be removed from the SAM in order to compare with the HOD mock catalogues. Finally in Section~4.5 we discuss the treatment of the radial distribution of satellite galaxies within haloes. 

%%%%%%%%%%%%%%%%%%%%%%%%%%%%%%%%%%%%%%%%%%%%%%%%%%%%%%%%%%%%%%%%%%%%%%%%%%%%%%%
\subsection{The HOD models used to build mocks}
\label{subsec: HOD models}

We test three different HOD schemes of increasing complexity. This helps us to understand the level of complexity needed to obtain accurate clustering predictions. Each model uses occupation functions obtained from linear interpolations of the HOD values in each bin, rather than fitting a parametric form to the SAM HOD. The distribution of galaxies can be nearest-integer (centrals only) and Poisson or negative binomial (satellites).

\subsubsection{1-HOD}

The 1-HOD model builds mock catalogues using the HOD of all galaxies from the SAM sample (black solid lines in Fig.~\ref{fig: HODs from SAMs}) including both centrals and satellites. The model assumes either a Poisson or negative binomial distribution for the occupation number. We adopt a Monte Carlo approach to obtain the final number of galaxies. 

This approach does not distinguish between centrals and satellites. If the model predicts that $\rm N \geq 1$ we assume that this halo hosts a central and $\rm N_{sat}=N-1$. Because of this, the number of centrals and satellites in the 1-HOD mock catalogues can be notably different with respect to the SAM samples where there are haloes with satellites but no central. Moreover, the HODs of these two separate components in the mock catalogues are completely different with respect to the HODs of the SAM samples (see Appendix~\ref{section:1-HODoccupationfunctions}). However, the total number of galaxies in these mock catalogues is essentially the same as in the SAM samples. 

\subsubsection{2-HOD}

The 2-HOD model uses the HOD of centrals and satellites separately, i.e, the red and blue solid lines in Fig.~\ref{fig: HODs from SAMs}, respectively. Thus, a particular distribution can be assumed for each component and the modelling is done independently for each one. For centrals, we use the nearest-integer distribution and for satellites the Poisson or negative binomial distribution. 

This scheme predicts practically the same number of central and satellites as the SAM samples. Note that in a non-negligible number of realizations it is possible to get haloes without a central. This is more likely for haloes with masses for which $\rm \left\langle N_{cen}\right\rangle <  1$ which is more frequently found in SFR-selected samples.

\subsubsection{4-HOD}

This model contains more information about the galaxy population of the SAM sample than the 2-HOD model. The 4-HOD requires us to store the number of haloes that host a central ($\rm N_{cen}$) and the number of haloes that do not host a central ($\rm N_{nocen}$) as a function of halo mass. Under this definition the total number of haloes in the volume is the sum of both quantities. Furthermore, the 4-HOD also needs knowledge of the number of satellites in haloes with a central ($\rm N_{sat\_cen}$) and without a central ($\rm N_{sat\_nocen}$). Thus, the total number of satellites is the sum of these two quantities. With these definitions, we build new HODs for satellites that take into account the population of centrals in the SAM samples. The SAM samples contain haloes with satellites but no centrals. This is more common in SFR selected samples. Indeed the HOD of centrals in these samples indicates that a large number of haloes do not host a central (see Fig~\ref{fig: HODs from SAMs}), and the 4-HOD takes this feature into account. 

We then define the satellite occupation functions conditioned on whether or not haloes host a central. With the four quantities explained above, we can define the conditional HODs,

\begin{align}
\rm \langle N_{\rm sat\_cen}(M_h) \rangle  &= \rm \frac{N_{sat\_cen}}{N_{cen}}(M_h) \label{eq. sat_cen} \\
\rm \langle N_{\rm sat\_nocen}(M_h) \rangle  &= \rm \frac{N_{sat\_nocen}}{N_{nocen}}(M_h) \label{eq. sat_nocen}
\end{align}

\begin{figure}
    \centering
    \includegraphics[width=\columnwidth]{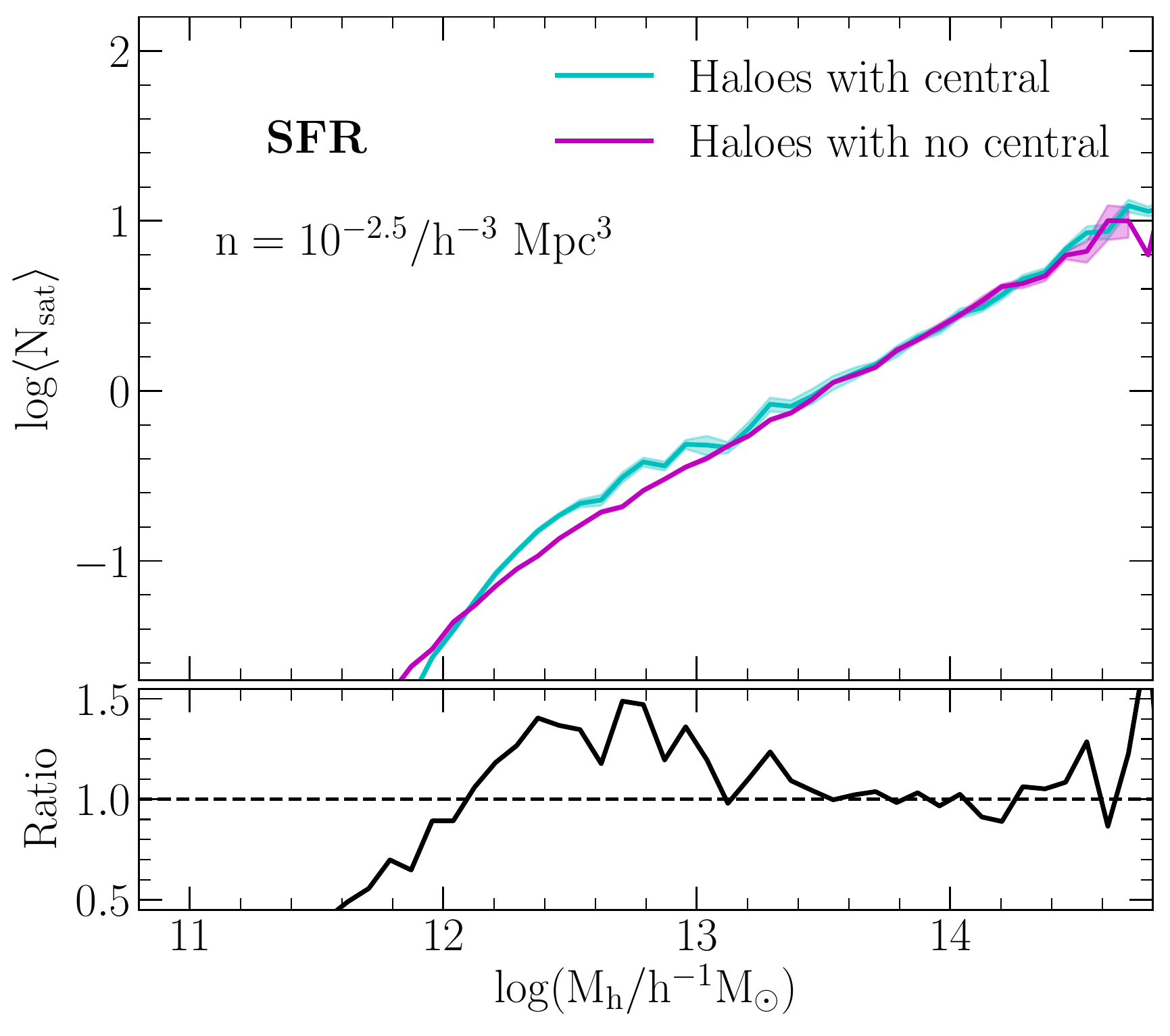}
    \caption{Conditional HODs from the 4-HOD method for an SFR selected sample with number density ${\rm 10^{-2.5}} h^3 \,{\rm Mpc^{-3}}$. {\it Top:} Average number of satellites in haloes with a central (cyan) and without a central (magenta). {\it Bottom:} Ratio of the two HODs shown in the upper panel. Shaded regions represent jackknife errors calculated using 10 subsamples.} 
    \label{fig: 4-HOD}
\end{figure}

Fig.~\ref{fig: 4-HOD} shows the conditional HODs where the main differences are observed at low halo masses. Even though the ratio between these two HODs is close to unity, it is the galaxies hosted by these haloes ($\approx 10^{12} h^{-1} {\rm M_{\odot}}$) that dominate the amplitude of clustering. The conditional HODs are well fitted by a negative binomial distribution, including the HODs of the other number density samples. The 4-HOD method uses a Monte Carlo approach to decide if a halo hosts a central galaxy. Depending on this outcome, one of the two conditional HODs is then chosen to obtain the number of satellites. 

%%%%%%%%%%%%%%%%%%%%%%%%%%%%%%%%%%%%%%%%%%%%%%%%%%%%%%%%%%%%%%%%%%%%%%%%%%%%
\subsection{Treatment of scatter in the HOD of satellites}
\label{subsec: scatter in the HOD}

\begin{figure}
	\centering
	\includegraphics[width=\columnwidth]{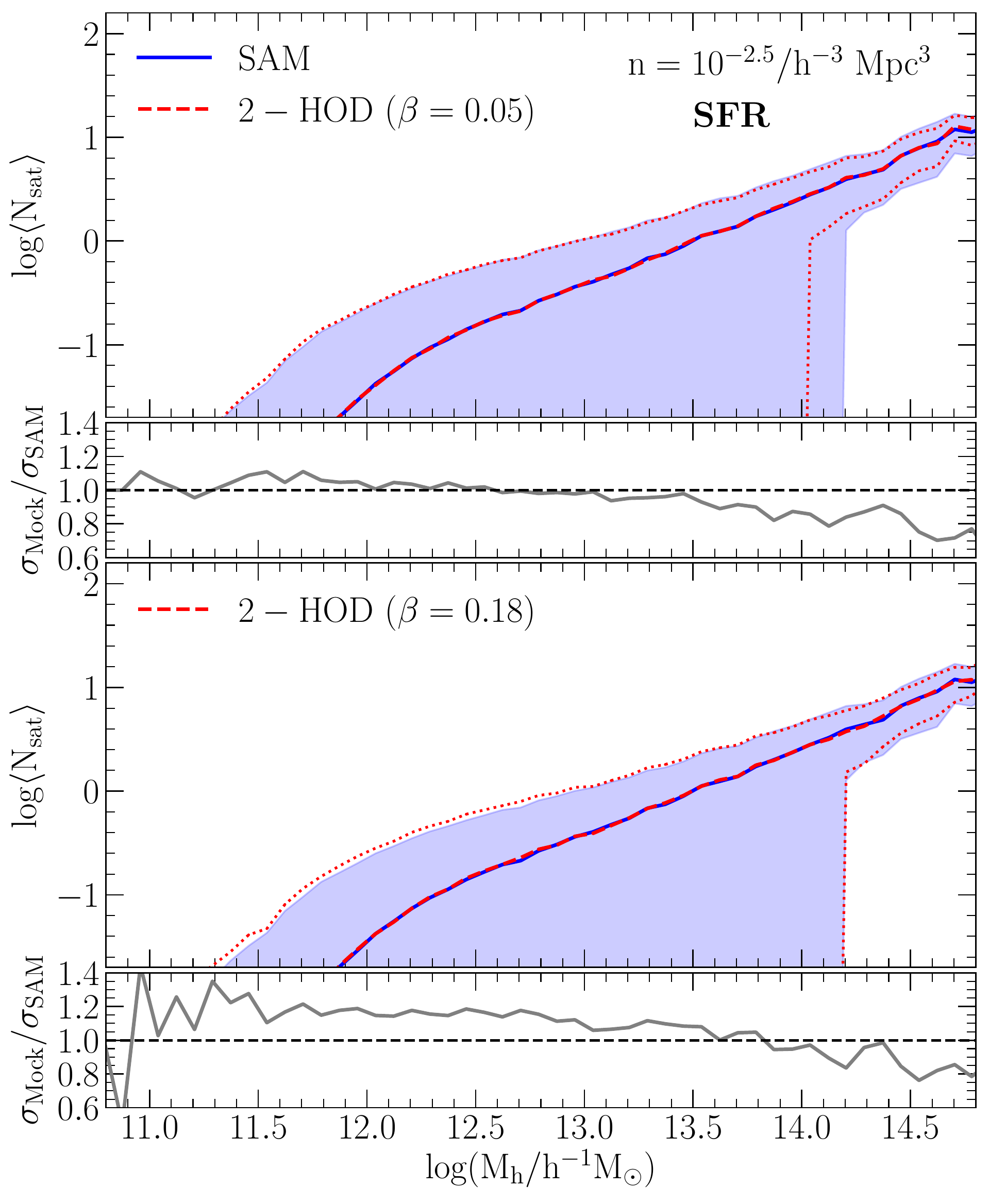}
    \caption{The HOD of satellites in a SAM sample (dashed blue) and a 2-HOD mock catalogue (solid red) contrasting two values of the parameter $\beta$ that controls the scatter (see Eq.~1): $\beta=0.05$ ({\it top}) and $\beta=0.18$ ({\it bottom}). The shaded regions show the HOD scatter  and the red dotted lines correspond to the  scatter in the HOD mocks. The subpanels show the ratios between the HOD scatter of the mocks and SAM sample. Note that it is not possible to visually distinguish a Poisson scatter from the $\beta$ scaled versions plotted in the main panels, but this choice would lead to a larger ratio of variances than the range plotted in the lower subpanels.} 
    \label{fig: beta definition}
\end{figure}

A Poisson distribution is fully described by its first moment. In the case of satellites this is $\rm \langle N_{sat} \rangle$.  If the distribution of the number of satellites follows instead a negative binomial distribution, an additional parameter $\beta$ is needed which specifies the increase in the scatter with respect to a Poisson distribution (see Eq.~1). We fix the $\beta$ value so that we reproduce as closely as possible the scatter of the HOD of satellites in a given SAM sample. Fig.~\ref{fig: beta definition} shows the scatter of the HOD of satellites in SAMs and 2-HOD mock catalogues for two illustrative $\beta$ values in a SFR selected sample. This shows that a small but non-zero $\beta$ is required to reproduce the HOD scatter of the SAM sample. The same is found for the other number density samples, and for the conditional HODs. We do not perform this analysis for the 1-HOD model as satellites are not treated independently in this case. 

It is not possible to replicate the HOD scatter in SAMs more closely as this would require $\beta$ to be a function of mass. Instead, we assume a constant scatter for the HOD of satellites by using the same $\beta$ for all halo mass bins. The accuracy of the $\beta$ values used are judged by checking the quality of the resulting mocks via comparison of their 2PCFs with the clustering of the shuffled-NFW samples (see \S~4.5 below for the definition of this catalogue). We show in Section~\ref{sec: results}, that when the scatter of the SAM and HOD mocks are matched up to  $M_h \lesssim 10^{13.5} M_{\odot}h^{-1}$ ($\beta=0.05$ for SFR-selected samples), we obtain the most accurate clustering predictions. In contrast, using larger values for $\beta$ worsens the predictions (as does using $\beta=0$, which corresponds to Poisson scatter). 

\begin{figure*}
  \begin{minipage}{\textwidth}
   \centering
      \includegraphics[width=0.8\textwidth]{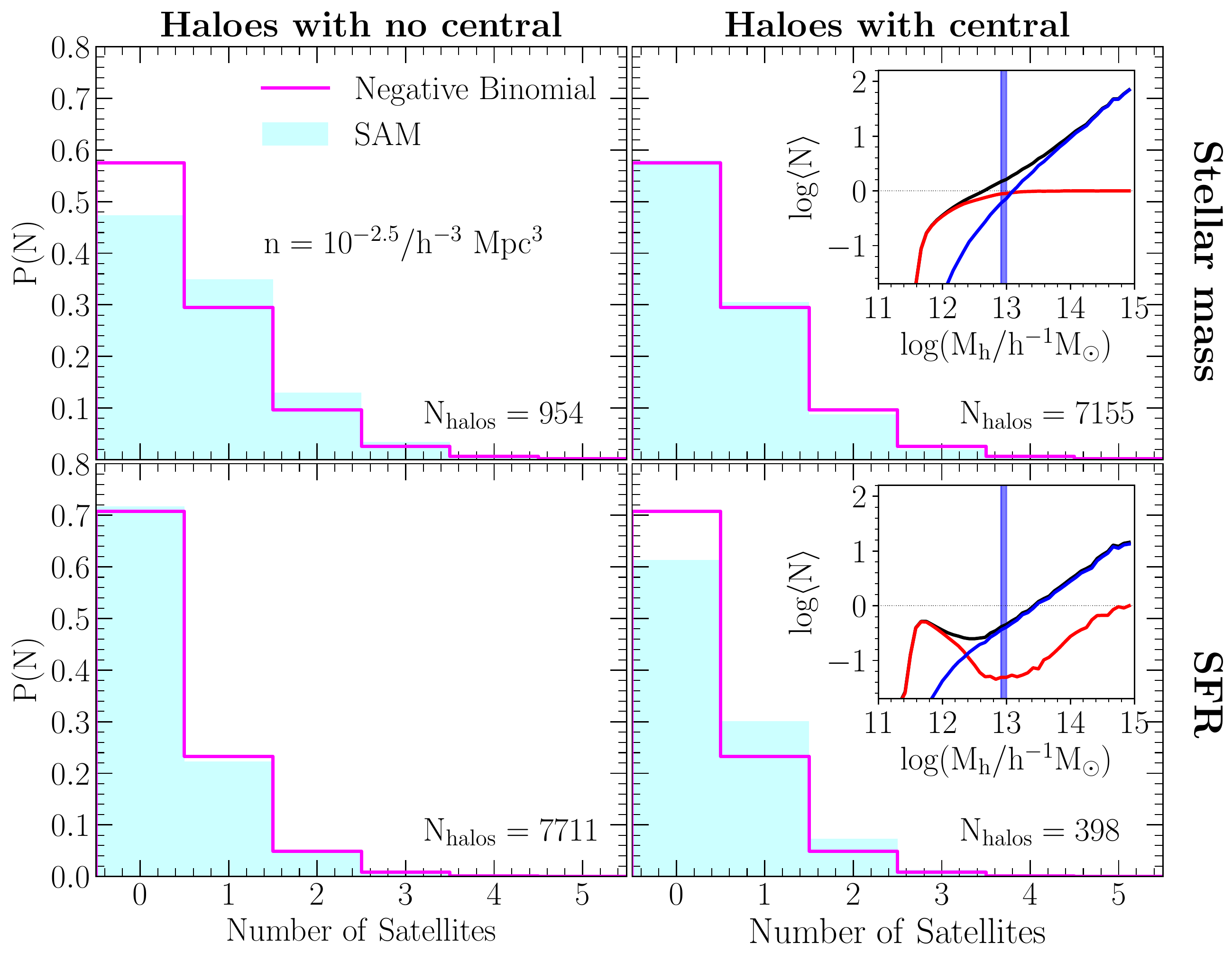}
      \caption{Probability distributions of satellites (cyan histograms) that are hosted by haloes with masses within the blue shaded (vertical) mass range  of the HODs shown in the insets. The galaxies are selected by stellar mass ({\it top}) and SFR ({\it bottom}), with a number density of ${\rm 10^{-2.5}} h^3 {\rm Mpc^{-3}}$. {\it Left:} Haloes without centrals (i.e its central did not enter to the cut). {\it Right} Same as left panel but considering haloes with centrals. Note the high probability to find haloes that do not host centrals in the SFR sample, which is expected by the low value of $\langle \rm  N_{cen} \rangle $ in the mass range analyzed and as is shown in the inset. The distributions are well described by negative binomial distributions (magenta) in the cases when most of haloes in the bin are included (i.e top right panel and bottom left panel). The negative binomial distributions shown here are obtained using a scatter that is $5\%$ larger than that from a Poisson distribution in the SFR selected sample, and $8\%$ larger in the stellar mass selection case}. 
     \label{fig: histograms}
  \end{minipage}
\end{figure*}

The satellite HOD is well described by the negative binomial distribution for a wide range of halo masses. Fig.~\ref{fig: histograms} shows the satellite PDF in a particular mass bin for a stellar mass and a SFR selected sample. We show negative binomial distributions defined by $\beta=0.08$ and $\beta=0.05$. In order to compute the satellite distributions, we split satellites according to whether or not their haloes host a central galaxy, which is relevant for the 4-HOD model. The satellite distribution matches with the negative binomial when most of haloes in the bin are included. A similar close match is found when comparing with Poisson distributions ($\beta=0$). Note that in the SFR selection case most of the haloes do not host a central galaxy, as the HOD of centrals in that bin suggests. The opposite behaviour is observed when selecting by stellar mass. 

%%%%%%%%%%%%%%%%%%%%%%%%%%%%%%%%%%%%%%%%%%%%%%%%%%%%%%%%%%%%%%%%%%%%%%%%%%%%
\subsection{The radial distribution of satellite galaxies in halos}
\label{subsec: distribution of satellites}

The number of satellites in the mock catalogues is obtained from the adopted HOD model (see Sec~\ref{subsec: HOD models}). Their positions in haloes are set according the standard NFW density profile \citep{Navarro96} which requires two parameters, the concentration and scale radius. The former depends on halo mass and the latter is a function of the virial radius. For simplicity, we assume that all haloes in the simulation volume have the same concentration parameter $c=13.98$ which corresponds to the concentration of a halo at redshift $\rm z=0$ and mass $M_h = 10^{12.5}{\rm M_{\odot}}h^{-1}$. We do not use a more realistic model for concentration as we are interested in comparing the HOD models rather than obtain a realistic redistribution of satellites. We impose that the maximum distance from a satellite to the halo center is two virial radii which depends on the halo mass. This defines the NFW mass profile used to obtain the satellite distances by a Monte Carlo approach. We modify the SAM output to impose a similar satellite distribution as described below (\S~4.5). 

%%%%%%%%%%%%%%%%%%%%%%%%%%%%%%%%%%%%%%%%%%%%%%%%%%%%%%%%%%%%%%%%%%%%%%%%%%%%%%%
\subsection{Removing assembly bias from the SAM output}
\label{subsec: AB and shuffle}

In order to determine the best methodology to produce HOD mock catalogues, we aim to compare them with the clustering of the original SAM samples via their 2PCF. Before making this comparison it is necessary to remove assembly bias from the SAM samples. 

The clustering of dark matter haloes depends on additional properties besides mass. For example, \citet{Gao05} showed that the clustering of low mass haloes depends on their formation redshift and other works have found dependencies on concentration and subhalo occupation number \citep[e.g.][]{Wechsler06} among other secondary properties. This additional contribution to the clustering is commonly known as assembly bias and potentially changes the galaxy clustering amplitude on large scales. 

The standard HOD approach considers only halo mass as the variable regulating the galaxy population. SAMs include assembly bias because they follow the evolution of baryons in halo merger histories that are shaped by the large-scale environment in the N-body simulation. Namely,  SAMs include a dependence on secondary halo properties as these affect the halo merger history and the evolution of galaxies that live within them. Thus, in order to compare the clustering between SAM samples and HOD mocks which use only halo mass as input, it is necessary to remove the assembly bias signal in the former samples. 

Assembly bias can be eliminated from SAM samples through the shuffling technique introduced by \citet{Croton07}. This consists of randomly exchanging the galaxy populations between haloes of the same mass, thus removing any connection to the assembly history of the haloes. This procedure does not change the distances from satellites to their central galaxy in each halo. In clustering terms, the one-halo term of this ``shuffled'' catalogue is the same as the original SAM sample but its two-halo term is different because assembly bias is not present in the shuffled sample. If the SAM samples did not have assembly bias, we would measure the same 2PCFs for their shuffled samples as measured for the original output. 

\begin{figure*}
 \begin{minipage}{\textwidth}
     \centering
     \includegraphics[width=0.95\textwidth]{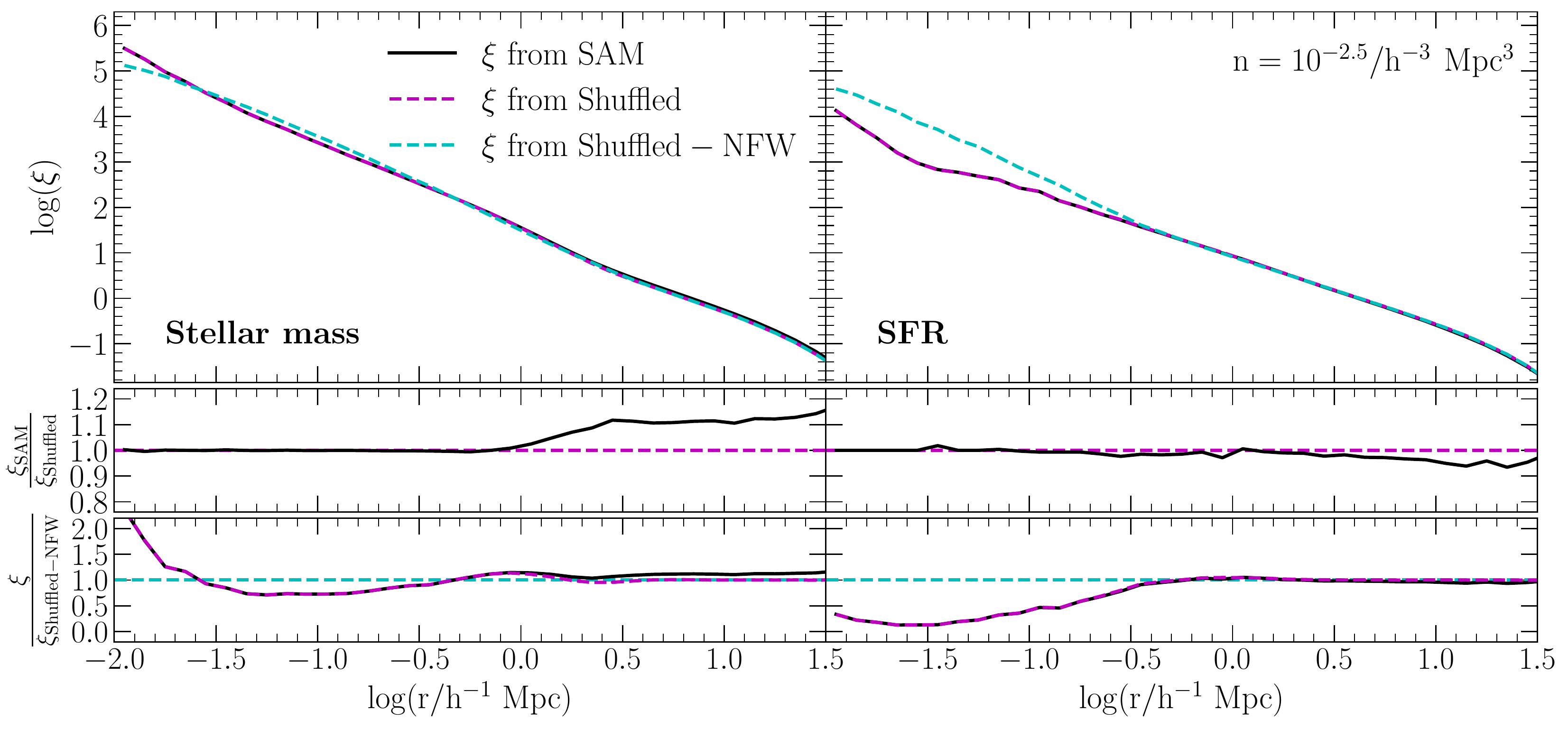}
    \caption{ {\it Top:} Correlation functions of SAM samples (dotted black in the main panel, solid black in the subpanel), and their modifications: the shuffled (dashed magenta) and the shuffled-NFW samples (dashed cyan). The galaxy samples are selected by stellar mass ({\it left}) and SFR ({\it right}), with a number density of ${\rm 10^{-2.5}} h^3{\rm Mpc^{-3}}$. {\it Middle:} Ratios between the 2PCF of the SAM and shuffled samples. Differences at large scales are signatures of assembly bias. {\it Bottom:} Ratios of the 2PCF with respect to the 2PCF from the shuffled-NFW samples. The differences in the one-halo term  below ${\rm 1\ Mpc/}h$ indicates the departure of the satellite profiles from a NFW.} 
    \label{fig: xi_SAM_Shuffle}
\end{minipage}
\end{figure*}

Fig.~\ref{fig: xi_SAM_Shuffle} shows the correlation functions of a SAM sample and its shuffled version, for both the stellar mass and the SFR selected samples. The assembly bias signature, shown in the middle panels, is evident in the clustering differences between these two catalogues at large separations. We also show the 2PCF of a modified shuffled sample that will be introduced below. The assembly bias signatures remain unchanged for the other samples, but they are noisier for the lowest number density samples as they contain fewer galaxies.

It can be seen that assembly bias increases the clustering for stellar mass selected samples, as was shown by \citet{Zehavi18}. SFR selected samples, on the other hand, show a decreased clustering amplitude. For the intermediate galaxy density sample, the assembly bias enhances the two-halo term by $\sim 12\%$ for the stellar mass selected sample and suppresses the amplitude in the SFR selection case by $\sim 4\%$. The enhance of clustering amplitude for the other stellar mass selections remains similar. For the SFR selections, we observe that suppression on clustering becomes weaker for higher density samples. Indeed, assembly bias can enhance the amplitude if the density of the sample if very high,  as shown in \citep{contreras19}.

\subsection{The shuffled-NFW target catalogue: changing the satellite distribution in the SAM}

The radial profile of satellites in the G13 SAM deviates from the standard NFW profile of dark matter within halos because the SAM associates galaxies with subhalos (or a proxy, such as the most bound particle, in the case of subhalos which are no longer resolved). The radial profile of subhalos is different from that of the dark matter (see e.g. \citealt{Angulo:2009}). The choice of which subhalos (and former subhalos) are associated with galaxies is driven by the galaxy formation model, which determines the luminosity of any galaxy associated with a subhalo and whether or not it has merged due to dynamical friction (only type 2 satellites, those that no longer have a resolved subhalo associated with them, are considered as candidates for galaxy mergers).  

The final step before testing the accuracy of the HOD models is to modify the shuffled SAM catalogue to force the satellites in each halo to follow an NFW profile.  We call the result the shuffled-NFW catalogue. Because satellite galaxies in the SAMs and shuffled samples do not follow an NFW profile, the one-halo term of their 2PCFs are different to the one-halo term of the shuffled-NFW sample, as shown in the bottom panel of Fig.~\ref{fig: xi_SAM_Shuffle}. The shuffled-NFW catalogue does not contain assembly bias, and the satellites follow the {\it same} NFW profile as adopted in the HOD mocks. We note that the shuffled-NFW is not intended to be the ``best'' prediction of galaxy clustering but rather is the target sample for the reconstructions using the HOD mocks which has a controlled 1-halo clustering pattern to facilitate testing.

%%%%%%%%%%%%%%%%%%%%%%%%%%%%%%%%%%%%%%%%%%%%%%%%%%%%%%%%%%%%%%%%%%%%%%%%%%%%%%%%%%%%%%%%
%                                         RESULTS
%%%%%%%%%%%%%%%%%%%%%%%%%%%%%%%%%%%%%%%%%%%%%%%%%%%%%%%%%%%%%%%%%%%%%%%%%%%%%%%%%%%%%%%%
\section{{\bf Testing the accuracy of the HOD models}}
\label{sec: results}

\subsection{Satellite radial distributions and clustering of HOD mocks}

\begin{figure}
    \centering
    \includegraphics[width=\columnwidth]{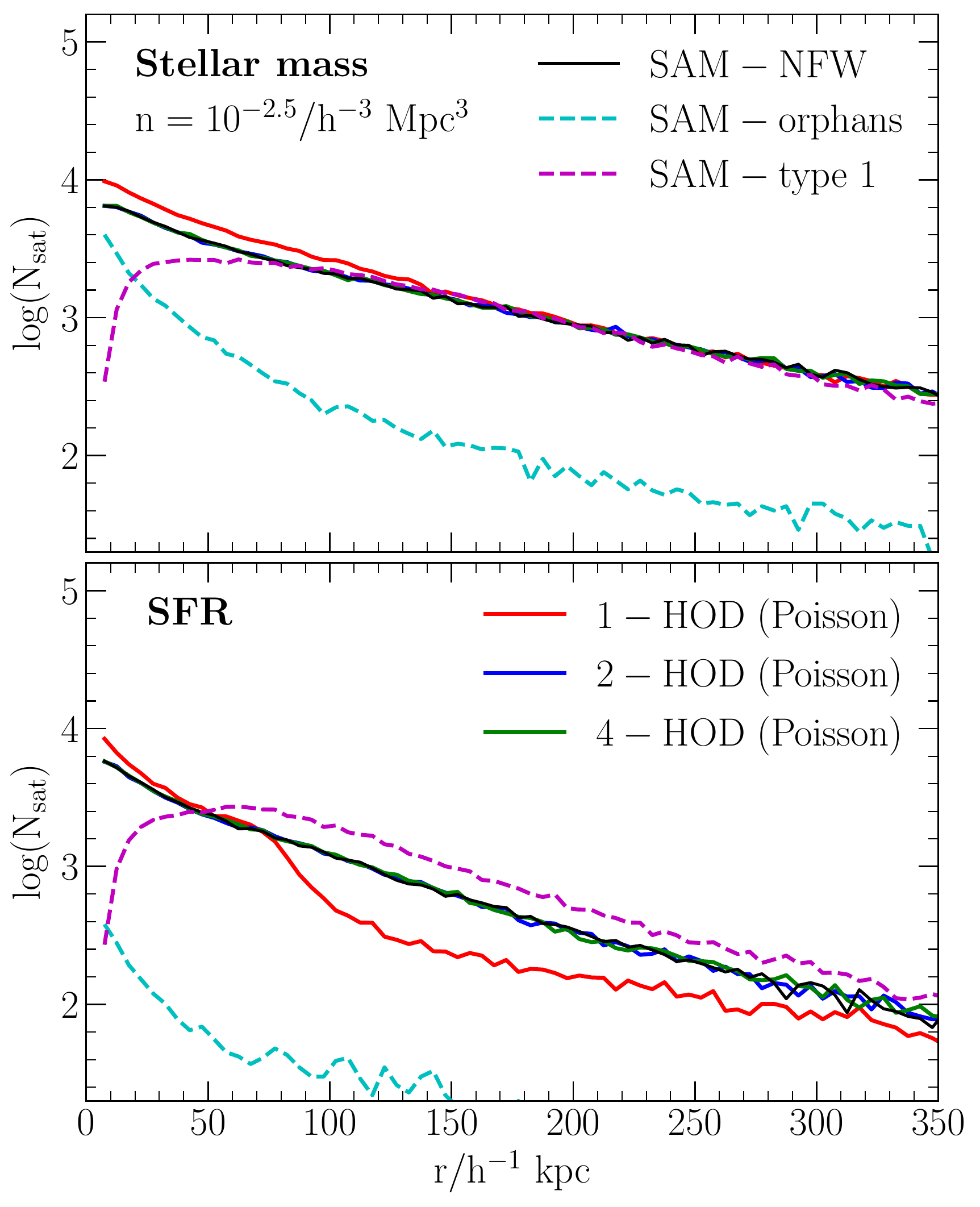}
    \caption{Profile of satellites hosted by subhaloes (dashed magenta) and orphans (dashed cyan) in a stellar mass ({\it top}) and SFR selected sample ({\it bottom}). The lines show the SAM with an NFW imposed for all satellites (solid black), and HOD mocks built by the 1-HOD (solid red), 2-HOD (solid blue) and 4-HOD models (solid green) where the HODs of satellites are described by the Poisson distribution.}
    \label{fig: rad_dist}
\end{figure}

In Fig.~\ref{fig: rad_dist} we show the satellite profiles in the SAM samples, for stellar mass and SFR selected samples separated into the contributions from type-1 satellites and from orphan galaxies. To examine the departure from NFW, we produce a SAM-NFW catalogue where satellites in SAM are forced to follow the same NFW profile as used in the HOD mocks. For this catalogue, we update the satellite positions in the SAM samples according the same NFW density profile used to produce the HOD mocks. Note that this SAM-NFW is different from the shuffled and shuffled-NFW catalogues mentioned above.

It can be seen that the NFW profile is different from  the profile of type-1 satellites and orphans, particularly for the SFR selected sample. The profiles in the 2-HOD and 4-HOD mock catalogues are also shown, and they match with the NFW as expected from the construction of the HOD mock. Both models also reproduce the NFW profile in the other number density samples.

The masses of host haloes of 1-HOD satellites do not correspond with the masses in the original SFR selected samples (see Fig.~\ref{fig: 1HOD results}). Thus, the virial radii of these haloes define NFW density profiles that are different from the profiles in the other models. This has an impact on the positions of satellites generating the striking difference with NFW in Fig.~\ref{fig: rad_dist} for the SFR selection. The same occurs for the stellar mass selections but it is less extreme than the SFR case.

The HOD models predict different galaxy populations for the G13 SAM samples. Table~\ref{table: sat_fractions} shows the satellite fraction of the SAM samples and HOD mocks built by the three different models, assuming a Poisson distribution for the HOD of satellites. Note that the 2-HOD and 4-HOD predict almost the same satellite fraction as in the SAM samples because of the separate HOD modelling of centrals and satellites.

\begin{table}
\caption{Satellite fractions of the galaxy samples used. The first column indicates their number densities. Column 2,3,4 and 5 show the satellite fraction in the SAM samples and in the HOD mock built using the 1-HOD, 2-HOD and 4-HOD models, respectively.} 
 \begin{tabular}{c c c c c } 
  \hline
 & & Stellar Mass & &  \\
 \hline
 $n/h^3 {\rm Mpc^{-3}}$  & SAM & 1-HOD & 2-HOD & 4-HOD \\ 
  $10^{-3.0}$ & $0.280$ & $0.317$ & $0.279$ & $0.280$ \\
  $10^{-2.5}$ & $0.322$ & $0.405$  & $0.324$ & $0.324$ \\
  $10^{-2.0}$ & $0.381$ & $0.517$ & $0.383$ & $0.383$\\
 \hline
  & & SFR & &  \\
 \hline
 $n/h^3 {\rm Mpc^{-3}}$ & SAM & 1-HOD & 2-HOD & 4-HOD \\ 
  $10^{-3.0}$ & $0.171$ & $0.084$ & $0.175$ & $0.172$ \\
  $10^{-2.5}$ & $0.195$ & $0.192$ & $0.197$ & $0.197$ \\
  $10^{-2.0}$ & $0.244$ & $0.334$ & $0.246$ & $0.246$ \\
  \hline
  \end{tabular}

\label{table: sat_fractions}
\end{table}

\begin{figure}
    \centering
    \includegraphics[width=\columnwidth]{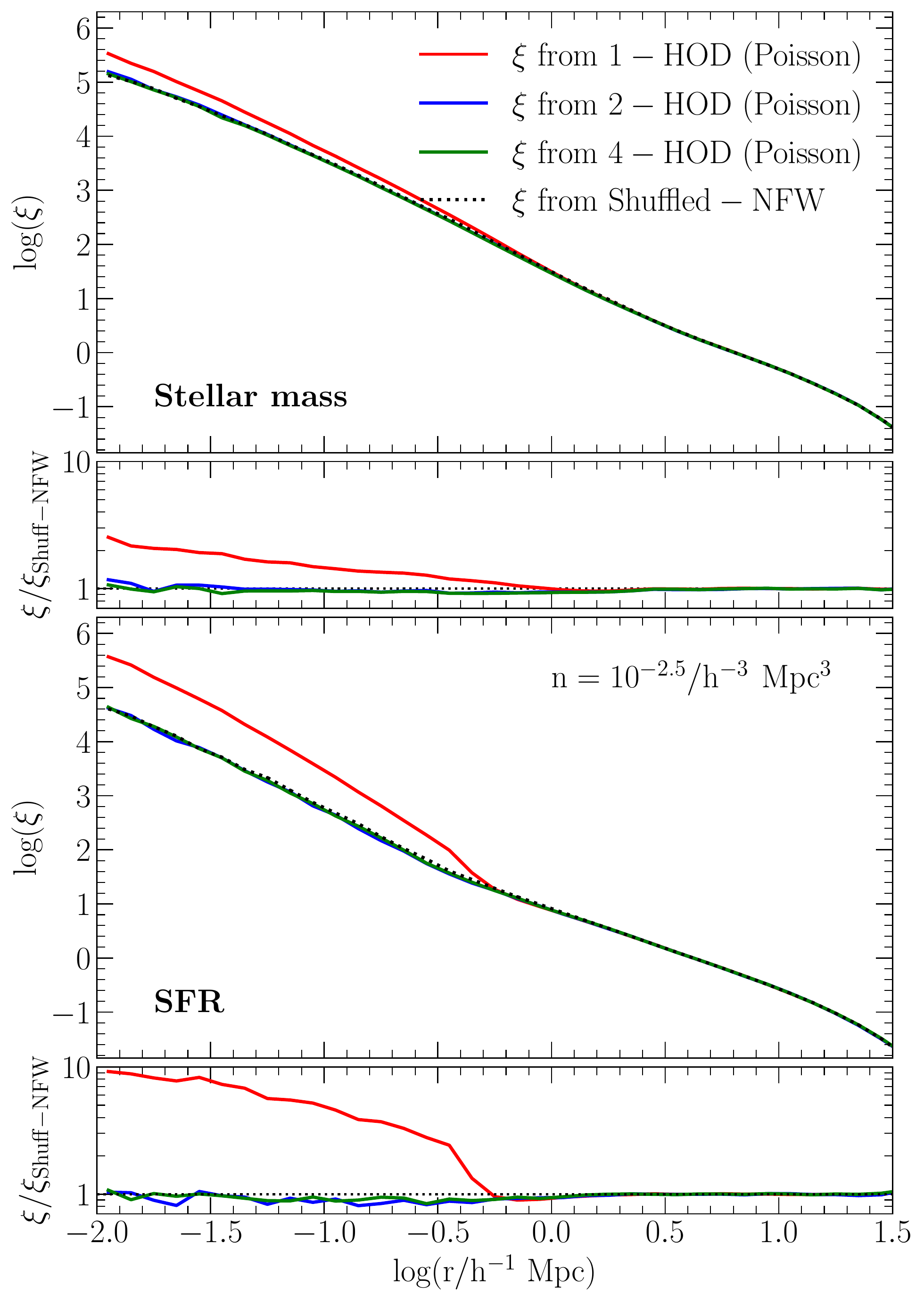}
    \caption{Clustering of HOD mock catalogues of stellar mass ({\it top}) and SFR selected samples ({\it bottom}) for a number density of ${\rm n=10^{-2.5}}/h^{-3}\rm Mpc^3$. The mocks are built using the 1-HOD (red), 2-HOD (blue) and 4-HOD (green) models, assuming a Poisson distribution for the HOD of satellites. The clustering of the shuffled catalogue with an NFW profile is shown as the dotted line. Subpanels show the ratios of the 2PCF of the mocks with respect to the 2PCF of the shuffle-NFW catalogue}
    \label{fig: xi_d3}
\end{figure}

We check the accuracy of each HOD model by comparing the HOD mocks with the shuffled-NFW sample via their 2PCFs. Fig.~\ref{fig: xi_d3} shows the clustering of the shuffled-NFW and the HOD mocks built using the HOD models described in Sec.~\ref{subsec: HOD models}. These particular models assume a Poisson distribution for the HOD of satellites. It can be seen that the three schemes produce accurate clustering predictions on large scales. On small scales, the 2-HOD and 4-HOD models produce similar accurate results while the 1-HOD shows striking differences. These deviations come from the overprediction of the number of satellites in the stellar mass selected samples. For the SFR selection cases, the difference is due to the notably different occupation function of central and satellites in the 1-HOD mock (see Appendix~\ref{section:1-HODoccupationfunctions}). The inaccuracy of the 1-HOD modelling is also present for the other number density samples too, whereas the 2-HOD and 4-HOD models produce similar quality results to the one shown here. As the 2-HOD and 4-HOD models are clearly the best, we drop the 1-HOD model henceforth.

%%%%%%%%%%%%%%%%%%%%%%%%%%%%%%%%%%%%%%%%%%%%%%%%%%%%%%%%%%%%%%%%%%%%%%%%%%%%

\subsection{Impact of the assumed HOD scatter}

\begin{figure*}
  \begin{minipage}{\textwidth}
      \centering
      \includegraphics[width=0.9\textwidth]{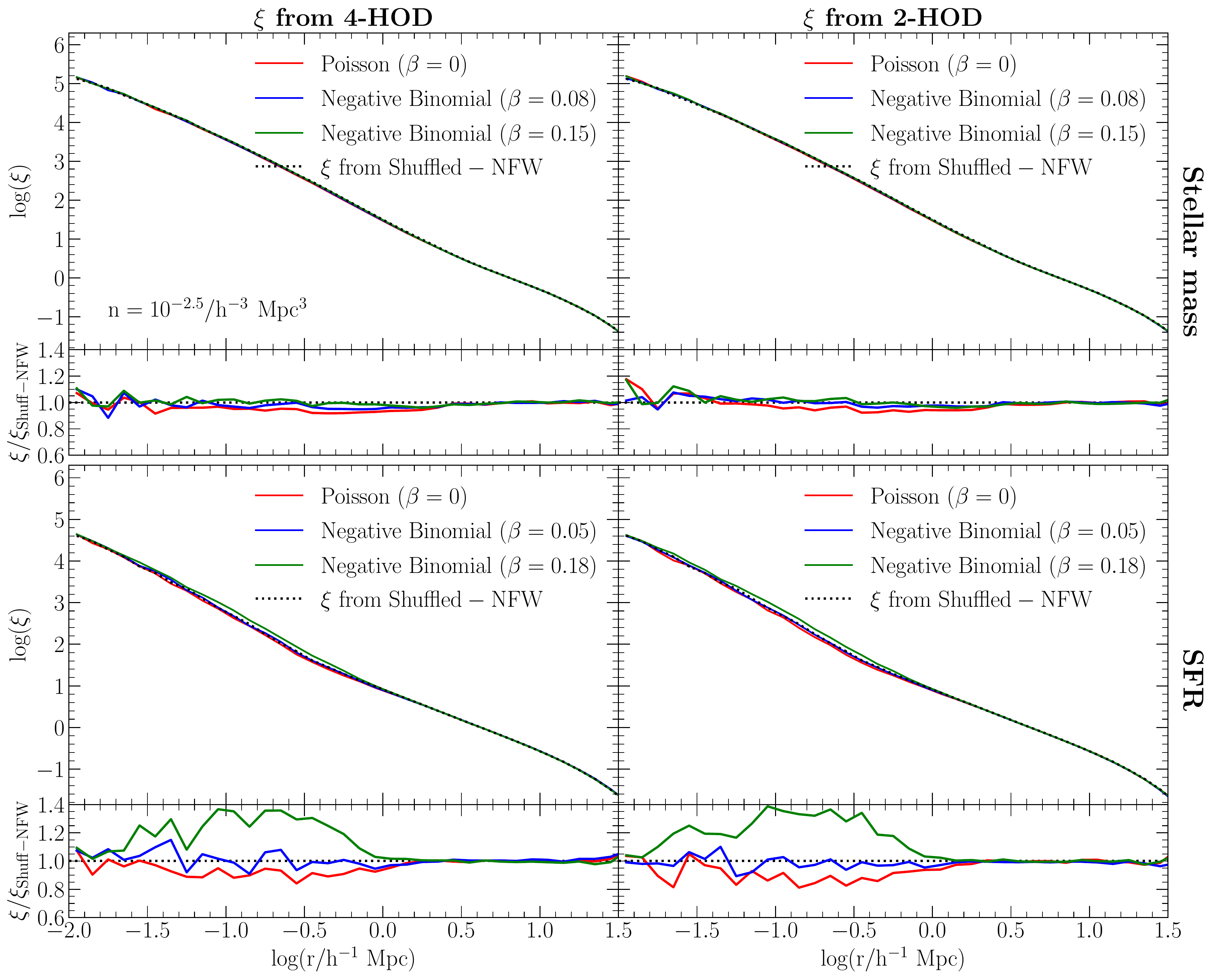}
      \caption{2PCF of HOD mocks built using the 2-HOD ({\it right}) and 4-HOD ({\it left}) models, for stellar mass ({\it top}) and SFR selected samples ({\it bottom}). The HOD models are used to build mock catalogues assuming a Poisson distribution (solid red) and negative binomial distributions of $\beta=0.05$ (solid blue) and $\beta=0.18$ (solid green) for the HOD of satellites. The clustering of the shuffled-NFW samples (dotted black) is shown in each panel. The ratios between the clustering of HOF mocks and shuffled-NFW are shown in the subpanels.}
      \label{fig: xi_bump}
  \end{minipage}
\end{figure*}

To study the impact of the scatter of the HOD on the clustering, we consider different dispersions for the negative binomial distribution in the construction of HOD mocks. Fig.~\ref{fig: xi_bump} shows the 2PCF of HOD mocks using different $\beta$ values. For the stellar mass selected samples, the scatter of the HOD does not have a significant impact. 

For the SFR selection, the amplitude of the clustering on small scales is very sensitive to the scatter in the number of satellites. We find that increasing $\beta$ changes the amplitude of the one-halo term. When we split the contribution to the clustering from low and high mass haloes, we observe that the scatter mainly impacts the one-halo term of low mass haloes. This feature is present in HOD mocks built using the 2-HOD and 4-HOD methods. This is reproduced by both HOD models, indicating that this is a feature particular to the SFR selected samples.

The most accurate clustering reconstructions for the G13 samples are obtained when we use the 2-HOD or 4-HOD to build mock catalogues assuming a negative binomial distribution for the HOD of satellites. Note that clustering predictions from both models do not show significant differences.

\begin{figure*}
  \begin{minipage}{\textwidth}
      \centering
      \includegraphics[width=0.9\textwidth]{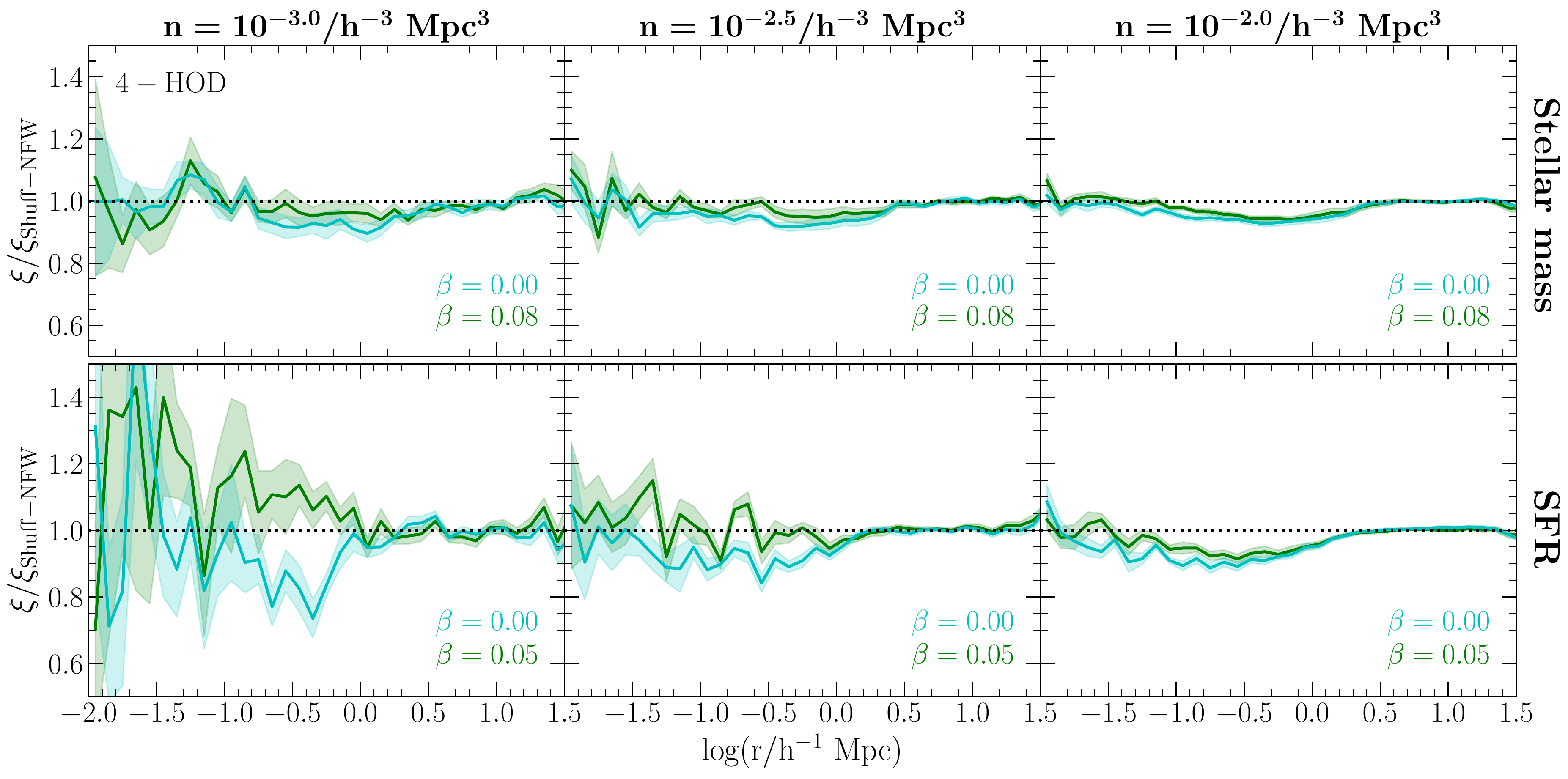}
      \caption{Ratios between the 2PCFs of mock catalogues, constructed with the 4-HOD method, and the shuffled-NFW catalogue. The HOD of satellites in the HOD mocks follows either the negative binomial (green) or Poisson distributions (cyan), with the color indicating the value of $\beta$. We show results for stellar mass ({\it top}) and SFR selected samples ({\it bottom}). Number densities increase from left to the right as labelled at the top of each column. The shaded regions represent jackknife errors calculated  using  10 subsamples.} 
     \label{fig: 4HOD_results}
  \end{minipage}
\end{figure*}

Fig~\ref{fig: 4HOD_results} shows the particular results from the 4-HOD modelling for all the space density samples. For the G13 SFR selected samples, the 4-HOD (and the 2-HOD) modelling produces the best results when $\beta=0.05$, which corresponds to a distribution slightly wider than a Poisson distribution. For the case of stellar mass selected samples, the best reproduction is obtained with $\beta=0.08$. Using instead the Poisson distribution (i.e, $\beta=0$) produces worse results for both selections particularly in the one halo regime. 

For SFR selections, when using $\beta=0.05$ and $\beta=0$, the departures from the shuffled-NFW catalogues are below $\sim 8\%$ and $\sim 15\%$, respectively. It can be seen that the dispersion of the 2PCFs becomes important in the lowest number density sample. However, the assumption of the negative binomial distribution still produces better results, especially in the transition from the one- to the two-halo term.

For the stellar mass selection cases, the impact on clustering when using different $\beta$ values is much less significant. Indeed, the weak relation between clustering and HOD scatter, shown in Fig.~\ref{fig: xi_bump}, suggests that it is not necessary to include additional scatter in the construction of HOD mock catalogues for stellar mass selections. To compare with the Poisson distribution, we show also the clustering prediction for the stellar mass samples using $\beta=0.08$, for which we obtain the best result.

Satellites in the G13 SAM are well described by a non-Poisson distribution. This is consistent with the HOD of subhaloes found in \citet{BK10}. The recipes that build mock catalogues of SFR selected samples using the HOD approach must undertake an analysis of the scatter of the HOD of satellites as it impacts the clustering. This analysis will provide the best $\beta$ to construct a HOD mock of a particular sample. For stellar mass samples, the HOD scatter has a weak impact on clustering, so the same analysis is not necessary in the context of HOD mock catalogues.

%%%%%%%%%%%%%%%%%%%%%%%%%%%%%%%%%%%%%%%%%%%%%%%%%%%%%%%%%%%%%%%%%%%%%%%%%%%%%%%%%%%%%%%%
%                                     CONCLUSIONS
%%%%%%%%%%%%%%%%%%%%%%%%%%%%%%%%%%%%%%%%%%%%%%%%%%%%%%%%%%%%%%%%%%%%%%%%%%%%%%%%%%%%%%%%
\section{Conclusion}
\label{sec: conclusions}

The next generation of surveys will measure the clustering of the galaxy distribution over a wide range of redshifts. Mock catalogues have proven to be important tools in preparation for this because of their multiple applications including error analysis, data interpretation and survey planning. SAMs are a physical approach to obtain such mocks, but 
sometimes their direct application to a simulation is not possible due to the limited resolution of the halo merger trees \citep{Angulo:2014} or the trees may not be available, as in the case of the Euclid flagship simulation \citep{Potter:2017}. Even if the trees were available at the required resolution, the sheer number of halos in a giga-parsec side N-body simulation may preclude a direct calculation with a SAM.

The HOD model provides a simple yet efficient way to construct mock catalogues. The modelling consists of using a probability distribution to obtain the number of galaxies hosted in a halo of a particular mass. This simple method allows us to create large sets of mock catalogues for huge cosmological simulations. This is useful in the context of ELGs, as they are targets in current and coming surveys.

To determine the level of complexity needed to produce accurate mock catalogues, we test different HOD models. The 1-HOD uses the HOD of all galaxies making no distinction between centrals and satellites while the 2-HOD uses the HOD of these two components separately. The 4-HOD stores additional information about whether or not haloes host a central, and it constructs conditional HODs for satellites taking into account this information.

Because SAMs include assembly bias by construction, and in their simplest form HOD mocks do not, we remove the assembly bias from the G13 SAM samples by shuffling the galaxy populations among haloes of the same mass creating the shuffled catalogue. This allows us to make a direct comparison between the clustering of our mocks and the SAMs from which we extract the HOD measurements. For example, We find that, for the intermediate galaxy density sample in the G13 SAM, the assembly bias affects the 2-halo term of the 2PCF of stellar mass selected galaxies increasing the amplitude by $\sim 12\%$. For the SFR selected galaxies, in contrast, the assembly bias suppresses the clustering by  $\sim 4\%$.  We also impose the standard NFW profile for satellites in the shuffled catalogue as is done for satellites in the HOD mock catalogues. We can then check the accuracy of the HOD models through a comparison between the 2PCFs of the HOD mocks and the shuffled-NFW catalogues.

The 2-HOD and 4-HOD produce the best mock catalogues as their 2PCFs are in close agreement with the clustering of the shuffled-NFW sample. We obtain the best results using a negative binomial distribution for the (conditional) HOD (see Eq.~1); in previous works this was commonly considered to be a Poisson distribution. This is consistent with the subhalo HOD found in \citet{BK10} using the Millennium-II simulation. Furthermore, we found that the assumption of this non-Poisonnian HOD changes the galaxy clustering. Previously, \citet{Jiang16} found a similar result using subhaloes from the Bolshoi, the MultiDark N-body simulations, and the SAM presented in \citet{Jian16-I}. 

The scatter of the HOD of satellites in G13 is reproduced by  a negative binomial distribution up to halo masses of ${\rm M_h \lesssim 10^{13.5}M_{\odot}}h^{-1}$. The galaxies in this halo mass range dominate the amplitude of the 2PCF. 
We quantify the departure from the Poisson distribution with the parameter $\beta$ (see Eq. \ref{eq. parameters of NB}). We obtain the best clustering predictions for SFR selected samples using $\beta=0.05$ and $\beta = 0.08$ for stellar mass selected samples. These correspond to negative binomial distributions slightly wider than Poisson. Because of the specific modelling of different SAMs, we expect that the best $\beta$ values for each sample are model-dependent. For stellar mass selected samples, we find that the HOD scatter has a weak impact on clustering, making the addition of this additional parameter unnecessary in the context of mock catalogues.

The analysis of the HOD of satellites is important because the width of the distribution (determined by the $\beta$ parameter) has a large impact on the one-halo term of the 2PCF of mock catalogues that emulate SFR-selected sample and ELG samples. If we consider the Poisson distribution for the HOD of satellites ($\beta=0$) the 2PCF of the mock catalogues is underestimated with respect to the clustering of the shuffled-NFW. In contrast, using the negative binomial distribution increases the amplitude of clustering in the one-halo regime. If we assume a value of $\beta$ larger than the one present in the distribution of number of satellites, the clustering on small scales is further overestimated. We highlight the importance to perform a careful analysis of the satellite HOD if the HOD framework is used to produce mock catalogues, for ELGs or star forming galaxies, following a particular model or observation.

\section*{Acknowledgements}

This work was made possible by the efforts of Gerard Lemson and colleagues at the German Astronomical Virtual Observatory in setting up the Millennium Simulation database in Garching. 
EJ, SC, NP \& IZ acknowledge the hospitality of the ICC at Durham University.
EJ acknowledges support from ``Centro de Astronom\'{i}a y Tecnolog\'{i}as Afines'' BASAL 170002. 
NP acknowledges support from Fondecyt Regular 1191813.
IZ acknowledges support by NSF grant AST-1612085.
This project has received funding from the European Union's Horizon 2020 Research and Innovation Programme
under the Marie Sk\l{}odowska-Curie grant agreement No 734374. The calculations for this paper were performed 
on the Geryon computer at the Center for Astro-Engineering UC, part of the BASAL PFB-06, which received additional
funding from QUIMAL 130008 and Fondequip AIC-57 for upgrades.

\bibliography{Biblio} 

\begin{thebibliography}{}
\makeatletter
\relax
\def\mn@urlcharsother{\let\do\@makeother \do\$\do\&\do\#\do\^\do\_\do\%\do\~}
\def\mn@doi{\begingroup\mn@urlcharsother \@ifnextchar [ {\mn@doi@}
  {\mn@doi@[]}}
\def\mn@doi@[#1]#2{\def\@tempa{#1}\ifx\@tempa\@empty \href
  {http://dx.doi.org/#2} {doi:#2}\else \href {http://dx.doi.org/#2} {#1}\fi
  \endgroup}
\def\mn@eprint#1#2{\mn@eprint@#1:#2::\@nil}
\def\mn@eprint@arXiv#1{\href {http://arxiv.org/abs/#1} {{\tt arXiv:#1}}}
\def\mn@eprint@dblp#1{\href {http://dblp.uni-trier.de/rec/bibtex/#1.xml}
  {dblp:#1}}
\def\mn@eprint@#1:#2:#3:#4\@nil{\def\@tempa {#1}\def\@tempb {#2}\def\@tempc
  {#3}\ifx \@tempc \@empty \let \@tempc \@tempb \let \@tempb \@tempa \fi \ifx
  \@tempb \@empty \def\@tempb {arXiv}\fi \@ifundefined
  {mn@eprint@\@tempb}{\@tempb:\@tempc}{\expandafter \expandafter \csname
  mn@eprint@\@tempb\endcsname \expandafter{\@tempc}}}

\bibitem[\protect\citeauthoryear{{Angulo}, {Lacey}, {Baugh}  \&
  {Frenk}}{{Angulo} et~al.}{2009}]{Angulo:2009}
{Angulo} R.~E.,  {Lacey} C.~G.,  {Baugh} C.~M.,   {Frenk} C.~S.,  2009, \mn@doi
  [\mnras] {10.1111/j.1365-2966.2009.15333.x}, \href
  {https://ui.adsabs.harvard.edu/abs/2009MNRAS.399..983A} {399, 983}

\bibitem[\protect\citeauthoryear{{Angulo}, {White}, {Springel}  \&
  {Henriques}}{{Angulo} et~al.}{2014}]{Angulo:2014}
{Angulo} R.~E.,  {White} S.~D.~M.,  {Springel} V.,   {Henriques} B.,  2014,
  \mn@doi [\mnras] {10.1093/mnras/stu905}, \href
  {https://ui.adsabs.harvard.edu/abs/2014MNRAS.442.2131A} {442, 2131}

\bibitem[\protect\citeauthoryear{{Baugh}}{{Baugh}}{2006}]{Baugh06-rv}
{Baugh} C.~M.,  2006, \mn@doi [Reports on Progress in Physics]
  {10.1088/0034-4885/69/12/R02}, \href
  {http://adsabs.harvard.edu/abs/2006RPPh...69.3101B} {69, 3101}

\bibitem[\protect\citeauthoryear{{Baugh}, {Benson}, {Cole}, {Frenk}  \&
  {Lacey}}{{Baugh} et~al.}{1999}]{Baugh:1999}
{Baugh} C.~M.,  {Benson} A.~J.,  {Cole} S.,  {Frenk} C.~S.,   {Lacey} C.~G.,
  1999, \mn@doi [\mnras] {10.1046/j.1365-8711.1999.02590.x}, \href
  {https://ui.adsabs.harvard.edu/abs/1999MNRAS.305L..21B} {305, L21}

\bibitem[\protect\citeauthoryear{{Benson}}{{Benson}}{2010}]{Benson10}
{Benson} A.~J.,  2010, \mn@doi [\physrep] {10.1016/j.physrep.2010.06.001},
  \href {http://adsabs.harvard.edu/abs/2010PhR...495...33B} {495, 33}

\bibitem[\protect\citeauthoryear{{Benson}, {Cole}, {Frenk}, {Baugh}  \&
  {Lacey}}{{Benson} et~al.}{2000}]{Benson00}
{Benson} A.~J.,  {Cole} S.,  {Frenk} C.~S.,  {Baugh} C.~M.,   {Lacey} C.~G.,
  2000, \mn@doi [\mnras] {10.1046/j.1365-8711.2000.03101.x}, \href
  {http://adsabs.harvard.edu/abs/2000MNRAS.311..793B} {311, 793}

\bibitem[\protect\citeauthoryear{{Berlind} \& {Weinberg}}{{Berlind} \&
  {Weinberg}}{2002}]{Berlind02}
{Berlind} A.~A.,  {Weinberg} D.~H.,  2002, \mn@doi [\apj] {10.1086/341469},
  \href {http://adsabs.harvard.edu/abs/2002ApJ...575..587B} {575, 587}

\bibitem[\protect\citeauthoryear{{Boylan-Kolchin}, {Springel}, {White},
  {Jenkins}  \& {Lemson}}{{Boylan-Kolchin} et~al.}{2009}]{BK09}
{Boylan-Kolchin} M.,  {Springel} V.,  {White} S.~D.~M.,  {Jenkins} A.,
  {Lemson} G.,  2009, \mn@doi [\mnras] {10.1111/j.1365-2966.2009.15191.x},
  \href {http://adsabs.harvard.edu/abs/2009MNRAS.398.1150B} {398, 1150}

\bibitem[\protect\citeauthoryear{{Boylan-Kolchin}, {Springel}, {White}  \&
  {Jenkins}}{{Boylan-Kolchin} et~al.}{2010}]{BK10}
{Boylan-Kolchin} M.,  {Springel} V.,  {White} S.~D.~M.,   {Jenkins} A.,  2010,
  \mn@doi [\mnras] {10.1111/j.1365-2966.2010.16774.x}, \href
  {http://adsabs.harvard.edu/abs/2010MNRAS.406..896B} {406, 896}

\bibitem[\protect\citeauthoryear{{Cochrane} \& {Best}}{{Cochrane} \&
  {Best}}{2018}]{Cochrane18}
{Cochrane} R.~K.,  {Best} P.~N.,  2018, \mn@doi [\mnras]
  {10.1093/mnras/sty1708}, \href
  {https://ui.adsabs.harvard.edu/abs/2018MNRAS.480..864C} {480, 864}

\bibitem[\protect\citeauthoryear{{Cochrane}, {Best}, {Sobral}, {Smail}, {Wake},
  {Stott}  \& {Geach}}{{Cochrane} et~al.}{2017}]{Cochrane17}
{Cochrane} R.~K.,  {Best} P.~N.,  {Sobral} D.,  {Smail} I.,  {Wake} D.~A.,
  {Stott} J.~P.,   {Geach} J.~E.,  2017, \mn@doi [\mnras]
  {10.1093/mnras/stx957}, \href
  {https://ui.adsabs.harvard.edu/abs/2017MNRAS.469.2913C} {469, 2913}

\bibitem[\protect\citeauthoryear{{Cole}, {Lacey}, {Baugh}  \& {Frenk}}{{Cole}
  et~al.}{2000}]{cole2000}
{Cole} S.,  {Lacey} C.~G.,  {Baugh} C.~M.,   {Frenk} C.~S.,  2000, \mn@doi
  [\mnras] {10.1046/j.1365-8711.2000.03879.x}, \href
  {http://adsabs.harvard.edu/abs/2000MNRAS.319..168C} {319, 168}

\bibitem[\protect\citeauthoryear{{Conroy}, {Wechsler}  \& {Kravtsov}}{{Conroy}
  et~al.}{2006}]{Conroy06}
{Conroy} C.,  {Wechsler} R.~H.,   {Kravtsov} A.~V.,  2006, \mn@doi [\apj]
  {10.1086/503602}, \href {http://adsabs.harvard.edu/abs/2006ApJ...647..201C}
  {647, 201}

\bibitem[\protect\citeauthoryear{{Contreras}, {Baugh}, {Norberg}  \&
  {Padilla}}{{Contreras} et~al.}{2013}]{contreras13}
{Contreras} S.,  {Baugh} C.~M.,  {Norberg} P.,   {Padilla} N.,  2013, \mn@doi
  [\mnras] {10.1093/mnras/stt629}, \href
  {http://adsabs.harvard.edu/abs/2013MNRAS.432.2717C} {432, 2717}

\bibitem[\protect\citeauthoryear{{Contreras}, {Zehavi}, {Padilla}, {Baugh},
  {Jim{\'e}nez}  \& {Lacerna}}{{Contreras} et~al.}{2019}]{contreras19}
{Contreras} S.,  {Zehavi} I.,  {Padilla} N.,  {Baugh} C.~M.,  {Jim{\'e}nez} E.,
    {Lacerna} I.,  2019, \mn@doi [\mnras] {10.1093/mnras/stz018}, \href
  {http://adsabs.harvard.edu/abs/2019MNRAS.484.1133C} {484, 1133}

\bibitem[\protect\citeauthoryear{{Croton} et~al.,}{{Croton}
  et~al.}{2006}]{Croton06}
{Croton} D.~J.,  et~al., 2006, \mn@doi [\mnras]
  {10.1111/j.1365-2966.2005.09675.x}, \href
  {http://adsabs.harvard.edu/abs/2006MNRAS.365...11C} {365, 11}

\bibitem[\protect\citeauthoryear{{Croton}, {Gao}  \& {White}}{{Croton}
  et~al.}{2007}]{Croton07}
{Croton} D.~J.,  {Gao} L.,   {White} S.~D.~M.,  2007, \mn@doi [\mnras]
  {10.1111/j.1365-2966.2006.11230.x}, \href
  {http://adsabs.harvard.edu/abs/2007MNRAS.374.1303C} {374, 1303}

\bibitem[\protect\citeauthoryear{{DESI Collaboration} et~al.,}{{DESI
  Collaboration} et~al.}{2016}]{DESI16}
{DESI Collaboration} et~al., 2016, arXiv e-prints, \href
  {https://ui.adsabs.harvard.edu/abs/2016arXiv161100036D} {p. arXiv:1611.00036}

\bibitem[\protect\citeauthoryear{{Davis}, {Efstathiou}, {Frenk}  \&
  {White}}{{Davis} et~al.}{1985}]{Davis85}
{Davis} M.,  {Efstathiou} G.,  {Frenk} C.~S.,   {White} S.~D.~M.,  1985,
  \mn@doi [\apj] {10.1086/163168}, \href
  {http://adsabs.harvard.edu/abs/1985ApJ...292..371D} {292, 371}

\bibitem[\protect\citeauthoryear{{De Lucia} \& {Blaizot}}{{De Lucia} \&
  {Blaizot}}{2007}]{DeLucia07}
{De Lucia} G.,  {Blaizot} J.,  2007, \mn@doi [\mnras]
  {10.1111/j.1365-2966.2006.11287.x}, \href
  {http://adsabs.harvard.edu/abs/2007MNRAS.375....2D} {375, 2}

\bibitem[\protect\citeauthoryear{{De Lucia}, {Kauffmann}  \& {White}}{{De
  Lucia} et~al.}{2004}]{DeLucia04}
{De Lucia} G.,  {Kauffmann} G.,   {White} S.~D.~M.,  2004, \mn@doi [\mnras]
  {10.1111/j.1365-2966.2004.07584.x}, \href
  {http://adsabs.harvard.edu/abs/2004MNRAS.349.1101D} {349, 1101}

\bibitem[\protect\citeauthoryear{{DeRose} et~al.,}{{DeRose}
  et~al.}{2019}]{DeRose19}
{DeRose} J.,  et~al., 2019, \mn@doi [\apj] {10.3847/1538-4357/ab1085}, \href
  {https://ui.adsabs.harvard.edu/abs/2019ApJ...875...69D} {875, 69}

\bibitem[\protect\citeauthoryear{{Gao}, {Springel}  \& {White}}{{Gao}
  et~al.}{2005}]{Gao05}
{Gao} L.,  {Springel} V.,   {White} S.~D.~M.,  2005, \mn@doi [\mnras]
  {10.1111/j.1745-3933.2005.00084.x}, \href
  {http://adsabs.harvard.edu/abs/2005MNRAS.363L..66G} {363, L66}

\bibitem[\protect\citeauthoryear{{Geach}, {Sobral}, {Hickox}, {Wake}, {Smail},
  {Best}, {Baugh}  \& {Stott}}{{Geach} et~al.}{2012}]{Geach12}
{Geach} J.~E.,  {Sobral} D.,  {Hickox} R.~C.,  {Wake} D.~A.,  {Smail} I.,
  {Best} P.~N.,  {Baugh} C.~M.,   {Stott} J.~P.,  2012, \mn@doi [\mnras]
  {10.1111/j.1365-2966.2012.21725.x}, \href
  {https://ui.adsabs.harvard.edu/abs/2012MNRAS.426..679G} {426, 679}

\bibitem[\protect\citeauthoryear{{Gonzalez-Perez} et~al.,}{{Gonzalez-Perez}
  et~al.}{2018}]{gp18}
{Gonzalez-Perez} V.,  et~al., 2018, \mn@doi [\mnras] {10.1093/mnras/stx2807},
  \href {http://adsabs.harvard.edu/abs/2018MNRAS.474.4024G} {474, 4024}

\bibitem[\protect\citeauthoryear{{Guo} et~al.,}{{Guo} et~al.}{2011}]{guo11}
{Guo} Q.,  et~al., 2011, \mn@doi [\mnras] {10.1111/j.1365-2966.2010.18114.x},
  \href {http://adsabs.harvard.edu/abs/2011MNRAS.413..101G} {413, 101}

\bibitem[\protect\citeauthoryear{{Guo}, {White}, {Angulo}, {Henriques},
  {Lemson}, {Boylan-Kolchin}, {Thomas}  \& {Short}}{{Guo} et~al.}{2013}]{Guo13}
{Guo} Q.,  {White} S.,  {Angulo} R.~E.,  {Henriques} B.,  {Lemson} G.,
  {Boylan-Kolchin} M.,  {Thomas} P.,   {Short} C.,  2013, \mn@doi [\mnras]
  {10.1093/mnras/sts115}, \href
  {http://adsabs.harvard.edu/abs/2013MNRAS.428.1351G} {428, 1351}

\bibitem[\protect\citeauthoryear{{Guo} et~al.,}{{Guo} et~al.}{2016}]{Guo16}
{Guo} Q.,  et~al., 2016, \mn@doi [\mnras] {10.1093/mnras/stw1525}, \href
  {https://ui.adsabs.harvard.edu/abs/2016MNRAS.461.3457G} {461, 3457}

\bibitem[\protect\citeauthoryear{{Henriques}, {White}, {Thomas}, {Angulo},
  {Guo}, {Lemson}  \& {Springel}}{{Henriques} et~al.}{2013}]{Henriques13}
{Henriques} B.~M.~B.,  {White} S.~D.~M.,  {Thomas} P.~A.,  {Angulo} R.~E.,
  {Guo} Q.,  {Lemson} G.,   {Springel} V.,  2013, \mn@doi [\mnras]
  {10.1093/mnras/stt415}, \href
  {http://adsabs.harvard.edu/abs/2013MNRAS.431.3373H} {431, 3373}

\bibitem[\protect\citeauthoryear{{Jiang} \& {van den Bosch}}{{Jiang} \& {van
  den Bosch}}{2016}]{Jian16-I}
{Jiang} F.,  {van den Bosch} F.~C.,  2016, \mn@doi [\mnras]
  {10.1093/mnras/stw439}, \href
  {https://ui.adsabs.harvard.edu/abs/2016MNRAS.458.2848J} {458, 2848}

\bibitem[\protect\citeauthoryear{{Jiang} \& {van den Bosch}}{{Jiang} \& {van
  den Bosch}}{2017}]{Jiang16}
{Jiang} F.,  {van den Bosch} F.~C.,  2017, \mn@doi [\mnras]
  {10.1093/mnras/stx1979}, \href
  {http://adsabs.harvard.edu/abs/2017MNRAS.472..657J} {472, 657}

\bibitem[\protect\citeauthoryear{{Klypin}, {Trujillo-Gomez}  \&
  {Primack}}{{Klypin} et~al.}{2011}]{Klypin11}
{Klypin} A.~A.,  {Trujillo-Gomez} S.,   {Primack} J.,  2011, \mn@doi [\apj]
  {10.1088/0004-637X/740/2/102}, \href
  {https://ui.adsabs.harvard.edu/abs/2011ApJ...740..102K} {740, 102}

\bibitem[\protect\citeauthoryear{{Kravtsov}, {Berlind}, {Wechsler}, {Klypin},
  {Gottl{\"o}ber}, {Allgood}  \& {Primack}}{{Kravtsov}
  et~al.}{2004}]{Kravtsov04}
{Kravtsov} A.~V.,  {Berlind} A.~A.,  {Wechsler} R.~H.,  {Klypin} A.~A.,
  {Gottl{\"o}ber} S.,  {Allgood} B.,   {Primack} J.~R.,  2004, \mn@doi [\apj]
  {10.1086/420959}, \href {http://adsabs.harvard.edu/abs/2004ApJ...609...35K}
  {609, 35}

\bibitem[\protect\citeauthoryear{{Laureijs} et~al.,}{{Laureijs}
  et~al.}{2011}]{Laureijs:2011}
{Laureijs} R.,  et~al., 2011, arXiv e-prints, \href
  {https://ui.adsabs.harvard.edu/abs/2011arXiv1110.3193L} {p. arXiv:1110.3193}

\bibitem[\protect\citeauthoryear{{Manera} et~al.,}{{Manera}
  et~al.}{2013}]{Manera13}
{Manera} M.,  et~al., 2013, \mn@doi [\mnras] {10.1093/mnras/sts084}, \href
  {http://adsabs.harvard.edu/abs/2013MNRAS.428.1036M} {428, 1036}

\bibitem[\protect\citeauthoryear{{Navarro}, {Frenk}  \& {White}}{{Navarro}
  et~al.}{1996}]{Navarro96}
{Navarro} J.~F.,  {Frenk} C.~S.,   {White} S.~D.~M.,  1996, \mn@doi [\apj]
  {10.1086/177173}, \href {http://adsabs.harvard.edu/abs/1996ApJ...462..563N}
  {462, 563}

\bibitem[\protect\citeauthoryear{{Norberg}, {Baugh}, {Gazta{\~n}aga}  \&
  {Croton}}{{Norberg} et~al.}{2009}]{Norberg:2009}
{Norberg} P.,  {Baugh} C.~M.,  {Gazta{\~n}aga} E.,   {Croton} D.~J.,  2009,
  \mn@doi [\mnras] {10.1111/j.1365-2966.2009.14389.x}, \href
  {https://ui.adsabs.harvard.edu/abs/2009MNRAS.396...19N} {396, 19}

\bibitem[\protect\citeauthoryear{{Orsi}, {Padilla}, {Groves}, {Cora}, {Tecce},
  {Gargiulo}  \& {Ruiz}}{{Orsi} et~al.}{2014}]{Orsi14}
{Orsi} {\'A}.,  {Padilla} N.,  {Groves} B.,  {Cora} S.,  {Tecce} T.,
  {Gargiulo} I.,   {Ruiz} A.,  2014, \mn@doi [\mnras] {10.1093/mnras/stu1203},
  \href {http://adsabs.harvard.edu/abs/2014MNRAS.443..799O} {443, 799}

\bibitem[\protect\citeauthoryear{{Peacock} \& {Smith}}{{Peacock} \&
  {Smith}}{2000}]{Peacokc00}
{Peacock} J.~A.,  {Smith} R.~E.,  2000, \mn@doi [\mnras]
  {10.1046/j.1365-8711.2000.03779.x}, \href
  {http://adsabs.harvard.edu/abs/2000MNRAS.318.1144P} {318, 1144}

\bibitem[\protect\citeauthoryear{{Potter}, {Stadel}  \& {Teyssier}}{{Potter}
  et~al.}{2017}]{Potter:2017}
{Potter} D.,  {Stadel} J.,   {Teyssier} R.,  2017, \mn@doi [Computational
  Astrophysics and Cosmology] {10.1186/s40668-017-0021-1}, \href
  {https://ui.adsabs.harvard.edu/abs/2017ComAC...4....2P} {4, 2}

\bibitem[\protect\citeauthoryear{{Prada}, {Klypin}, {Cuesta}, {Betancort-Rijo}
  \& {Primack}}{{Prada} et~al.}{2012}]{Prada12}
{Prada} F.,  {Klypin} A.~A.,  {Cuesta} A.~J.,  {Betancort-Rijo} J.~E.,
  {Primack} J.,  2012, \mn@doi [\mnras] {10.1111/j.1365-2966.2012.21007.x},
  \href {https://ui.adsabs.harvard.edu/abs/2012MNRAS.423.3018P} {423, 3018}

\bibitem[\protect\citeauthoryear{{Schaye} et~al.,}{{Schaye}
  et~al.}{2015}]{Schaye15}
{Schaye} J.,  et~al., 2015, \mn@doi [\mnras] {10.1093/mnras/stu2058}, \href
  {http://adsabs.harvard.edu/abs/2015MNRAS.446..521S} {446, 521}

\bibitem[\protect\citeauthoryear{{Scoccimarro}, {Feldman}, {Fry}  \&
  {Frieman}}{{Scoccimarro} et~al.}{2001}]{Scoccimarro01}
{Scoccimarro} R.,  {Feldman} H.~A.,  {Fry} J.~N.,   {Frieman} J.~A.,  2001,
  \mn@doi [\apj] {10.1086/318284}, \href
  {http://adsabs.harvard.edu/abs/2001ApJ...546..652S} {546, 652}

\bibitem[\protect\citeauthoryear{{Sinha} \& {Garrison}}{{Sinha} \&
  {Garrison}}{2017}]{Sinha17}
{Sinha} M.,  {Garrison} L.,  2017, {Corrfunc: Blazing fast correlation
  functions on the CPU}, Astrophysics Source Code Library (\mn@eprint {ascl}
  {1703.003})

\bibitem[\protect\citeauthoryear{{Somerville} \& {Dav{\'e}}}{{Somerville} \&
  {Dav{\'e}}}{2015}]{Sommerville15}
{Somerville} R.~S.,  {Dav{\'e}} R.,  2015, \mn@doi [\araa]
  {10.1146/annurev-astro-082812-140951}, \href
  {http://adsabs.harvard.edu/abs/2015ARA%26A..53...51S} {53, 51}

\bibitem[\protect\citeauthoryear{{Springel}, {White}, {Tormen}  \&
  {Kauffmann}}{{Springel} et~al.}{2001}]{Springel01}
{Springel} V.,  {White} S.~D.~M.,  {Tormen} G.,   {Kauffmann} G.,  2001,
  \mn@doi [\mnras] {10.1046/j.1365-8711.2001.04912.x}, \href
  {http://adsabs.harvard.edu/abs/2001MNRAS.328..726S} {328, 726}

\bibitem[\protect\citeauthoryear{{Springel} et~al.,}{{Springel}
  et~al.}{2005}]{Springel05}
{Springel} V.,  et~al., 2005, \mn@doi [\nat] {10.1038/nature03597}, \href
  {http://adsabs.harvard.edu/abs/2005Natur.435..629S} {435, 629}

\bibitem[\protect\citeauthoryear{{Vogelsberger} et~al.,}{{Vogelsberger}
  et~al.}{2014}]{Vogelsberger14}
{Vogelsberger} M.,  et~al., 2014, \mn@doi [\mnras] {10.1093/mnras/stu1536},
  \href {http://adsabs.harvard.edu/abs/2014MNRAS.444.1518V} {444, 1518}

\bibitem[\protect\citeauthoryear{{Wechsler} \& {Tinker}}{{Wechsler} \&
  {Tinker}}{2018}]{Wechsler18}
{Wechsler} R.~H.,  {Tinker} J.~L.,  2018, preprint, \href
  {http://adsabs.harvard.edu/abs/2018arXiv180403097W} {} (\mn@eprint {arXiv}
  {1804.03097})

\bibitem[\protect\citeauthoryear{{Wechsler}, {Zentner}, {Bullock}, {Kravtsov}
  \& {Allgood}}{{Wechsler} et~al.}{2006}]{Wechsler06}
{Wechsler} R.~H.,  {Zentner} A.~R.,  {Bullock} J.~S.,  {Kravtsov} A.~V.,
  {Allgood} B.,  2006, \mn@doi [\apj] {10.1086/507120}, \href
  {http://adsabs.harvard.edu/abs/2006ApJ...652...71W} {652, 71}

\bibitem[\protect\citeauthoryear{{White} \& {Rees}}{{White} \&
  {Rees}}{1978}]{White78}
{White} S.~D.~M.,  {Rees} M.~J.,  1978, \mn@doi [\mnras]
  {10.1093/mnras/183.3.341}, \href
  {http://adsabs.harvard.edu/abs/1978MNRAS.183..341W} {183, 341}

\bibitem[\protect\citeauthoryear{{Yang}, {Mo}  \& {van den Bosch}}{{Yang}
  et~al.}{2003}]{Yang03}
{Yang} X.,  {Mo} H.~J.,   {van den Bosch} F.~C.,  2003, \mn@doi [\mnras]
  {10.1046/j.1365-8711.2003.06254.x}, \href
  {http://adsabs.harvard.edu/abs/2003MNRAS.339.1057Y} {339, 1057}

\bibitem[\protect\citeauthoryear{{Zehavi} et~al.,}{{Zehavi}
  et~al.}{2011}]{Zehavi11}
{Zehavi} I.,  et~al., 2011, \mn@doi [\apj] {10.1088/0004-637X/736/1/59}, \href
  {http://adsabs.harvard.edu/abs/2011ApJ...736...59Z} {736, 59}

\bibitem[\protect\citeauthoryear{{Zehavi}, {Contreras}, {Padilla}, {Smith},
  {Baugh}  \& {Norberg}}{{Zehavi} et~al.}{2018}]{Zehavi18}
{Zehavi} I.,  {Contreras} S.,  {Padilla} N.,  {Smith} N.~J.,  {Baugh} C.~M.,
  {Norberg} P.,  2018, \mn@doi [\apj] {10.3847/1538-4357/aaa54a}, \href
  {http://adsabs.harvard.edu/abs/2018ApJ...853...84Z} {853, 84}

\bibitem[\protect\citeauthoryear{{Zheng} et~al.,}{{Zheng}
  et~al.}{2005}]{Zheng05}
{Zheng} Z.,  et~al., 2005, \mn@doi [\apj] {10.1086/466510}, \href
  {http://adsabs.harvard.edu/abs/2005ApJ...633..791Z} {633, 791}

\makeatother
\end{thebibliography}
\bibliographystyle{mnras}

%%%%%%%%%%%%%%%%%%%%%%%%%%%%%%%%%%%%%%%%%%%%%%%%%%

%%%%%%%%%%%%%%%%% APPENDICES %%%%%%%%%%%%%%%%%%%%%

\appendix

\section{1-HOD occupation functions}
\label{section:1-HODoccupationfunctions}

The 1-HOD model uses the HOD of all SAM galaxies in a sample to produce mock catalogues. This does not necessarily reproduce the HOD of central and satellites separately. 

\begin{figure}
	\centering
	\includegraphics[width=\columnwidth]{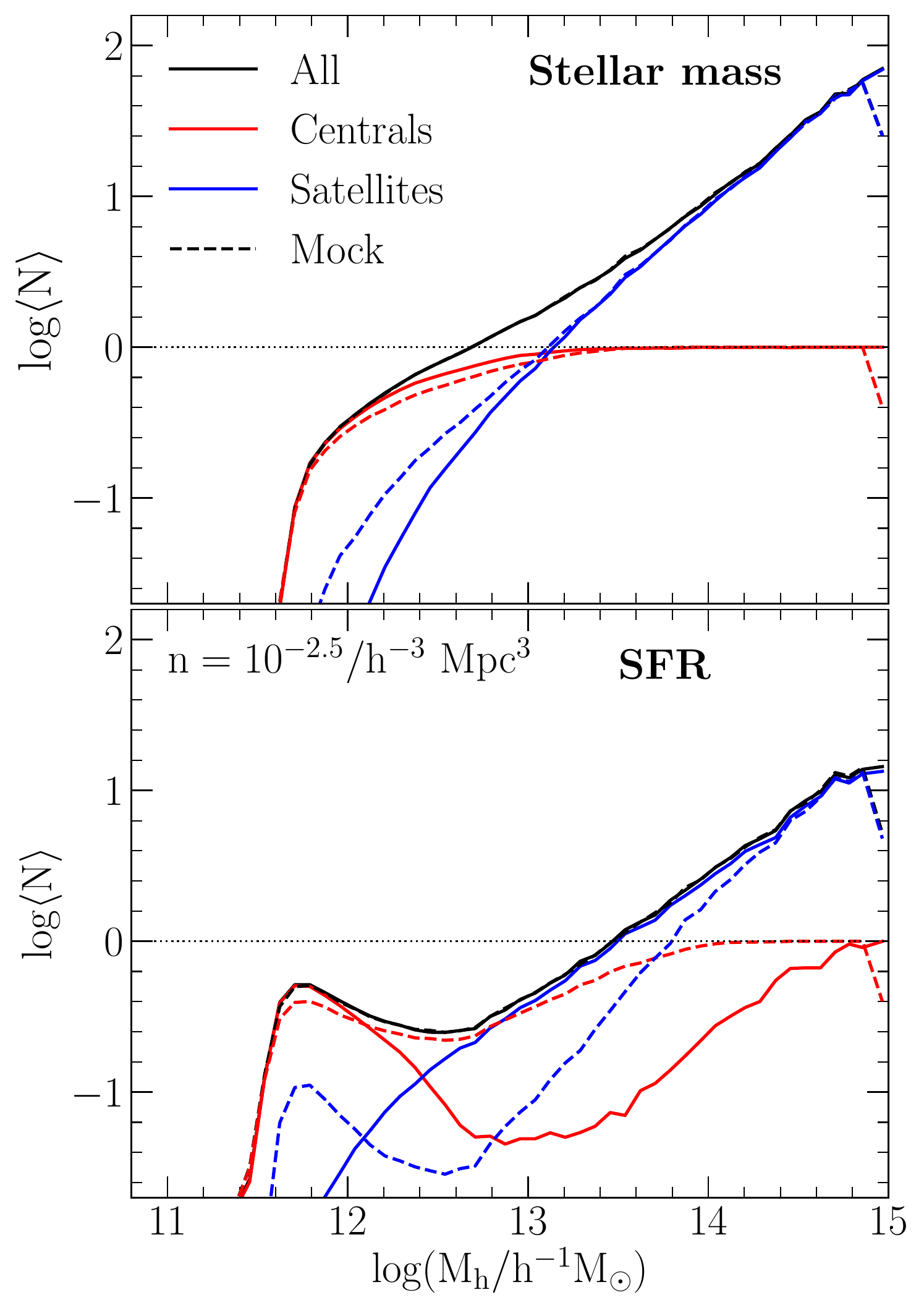}
    \caption{Same as Fig.~\ref{fig: HODs from SAMs} with the addition of the HODs of the 1-HOD mock catalogues  (dashed lines). Solid lines correspond to the HOD of SAM samples. The 1-HOD model reproduces the HOD of all galaxies but not the HOD of central and satellites separately.} 
    \label{fig: 1HOD results}
\end{figure}

The 1-HOD assumes a distribution for the full HOD, which we take to be Poisson or a negative binomial. However, centrals follow the nearest integer distribution, whereas satellites follow a Poisson or negative binomial distribution. As this distinction is not made, the predicted HODs of centrals and satellites in the resulting mock are different with respect to the original SAM samples.  Fig.~\ref{fig: 1HOD results} shows that 1-HOD tends to put satellites in very low mass haloes. Moreover, the occupation function of centrals in the SFR selection is overpredicted over a wide halo mass range.  

%\bsp	% typesetting comment
\label{lastpage}
\end{document}